\documentclass[aps,reprint,superscriptaddress,amsmath,amssymb,pra,floatfix]{revtex4-2}
\usepackage{physics}
\usepackage{siunitx}
\usepackage[colorlinks,linkcolor=blue,citecolor=blue,anchorcolor=blue,urlcolor=blue]{hyperref}
\usepackage{bm}
\usepackage{graphicx}
\usepackage{float}
\usepackage{xcolor}

\begin{document}

\preprint{APS/123-QED}

\title{Imaging moving atoms by holographically reconstructing the dragged slow light}

\author{Yuzhuo Wang$^1$}
\email{zhuodashi@163.com}

\author{Jian Zhao$^1$}
\email{Current address: KLA-Tencor Semiconductor Equipment Technology (Shanghai) Co., Ltd.}
\author{Xing Huang$^1$}
\author{Liyang Qiu$^1$}
\author{Lingjing Ji$^1$}
\author{Yudi Ma$^1$}
\author{Yizun He$^1$}
\author{James P. Sobol$^2$}
\author{Saijun Wu$^1$}
\email{saijunwu@fudan.edu.cn}

\affiliation{$^1$Department of Physics, State Key Laboratory of Surface Physics and Key
Laboratory of Micro and Nano Photonic Structures (Ministry of Education),
Fudan University, Shanghai 200433, China.\\
$^2$Keit Spectrometers, Harwell Campus, Oxfordshire UK.}

\begin{abstract}
    The propagation of light in moving media is dragged by atomic motion. The light-drag effect can be dramatically enhanced by reducing the group velocity with electro-magnetically induced transparency (EIT). We develop a systematic procedure to accurately reconstruct the complex wavefront of the slow light with single-shot measurements, enabling precise, photon shot-noise limited spectroscopic measurements of atomic response across EIT even in presence of generic atomic number fluctuations. Applying the technique to an expanding cloud of cold atoms, we demonstrate simultaneous inference of the atomic density distribution and the velocity field from the complex imaging data.  This inline imaging technique may assist a wide range of cold atom experiments to access spectroscopic and phase space information with in situ and minimally destructive measurements.
  \end{abstract}

 \maketitle
 
 \section{Introduction}
Investigations of light-drag effects in moving media have historically inspired people to better understand the nature of light, and for the establishment of special relativity~\cite{EINSTEIN1905,Gordon1923}. More recently, with the advance of quantum optical and cold atomic technology, there is resurgent research interest in light-drag effects~\cite{Strekalov2004,Safari2016,Kuan2016, Chen2020, Solomons2020}, particularly in coherently prepared optical media featuring electro-magnetically induced transparency (EIT)~\cite{Fleischhauer2005,Novikova2012}, where the light-drag effect is dramatically enhanced by the highly dispersive refractive index and enormously reduced group velocity~\cite{VestergaardHau1999}. The non-uniformly dragged dynamics of slow light in such medium have inspired exciting concepts and developments on optical analogy of gravitational effects~\cite{Leonhardt2002,Rosenberg2020}. Practically, since the dragged optical phase senses the relative motion between atoms and light, the effect is useful for inertial sensing and to support quantum enhanced performance of light pulse atomic interferometers~\cite{Matsko2003, Zimmer2004, Haine2015,Kuan2016,Chen2020}.

Previous studies of light-drag effects in atomic medium usually resort to optical interferometry for the phase readout~\cite{Matsko2003,Strekalov2004, Zimmer2004, Kuan2016,Chen2020}. The atomic sample is typically placed within one arm of the interferometer. The transmitted wavefront of the probe arm is compared with light in the reference arm at the output beamsplitter to measure the dragged phase shift. Although a single mode optical interferometer is ideal for sensing the average motion, spatial dependent information is disregarded in such setups. Visualization of the spatially dependent light-drag effect can form a powerful tool to retrieve phase-space information of dilute atomic samples with generic fluctuation of density and coupling strength to light. The phase space information may help, for example, to suppress classical noises in atom interferometers~\cite{Matsko2003, Dickerson2013} or to assist quantum feedback in degenerate gases~\cite{Engels2003,Wright2013,Galitski2019}. However, such visualization requires one to recover the complex wavefront of the slow light within single shot camera exposure, at sensitivity close to the photon shot-noise limit~\cite{Sobol2014} so as to minimize the photon-recoil induced back-actions. The required wavefront sensing capacity can in principle be achieved with holographic imaging by recovering optical phase information from intensity measurements~\cite{Cuche1999,2004microelectromechanical, Bjrn2005Investigation,2008Phase,Greenbaum2012,DeHaan2020}. To this end, efforts have been made over years for holographic imaging of cold atoms~\cite{Kadlecek2001,Turner2005,Sobol2014,Smits2020,Altuntas2021}. However, the achievable imaging accuracies in these work are limited by various approximations, typically designed to resolve the ``twin image problem'' associated with the phase ambiguity~\cite{Gabor1971}, such as by pre-assuming optical properties of atoms~\cite{Turner2005,Altuntas2021} or idealizing probe/reference wavefronts~\cite{Kadlecek2001,Turner2005,Sobol2014,Smits2020,Altuntas2021}. To suppress the associated errors is prerequisite to spatially resolving the kHz-level 2-photon Doppler shift of a dilute gas. The resulting atomic density-error immune spectroscopic imaging capacity has not been demonstrated previously, to the best of our knowledge. 

This work reports holographic reconstruction of slow light and spectroscopic imaging of moving atoms under the EIT condition. For the purpose, we develop an accurate procedure to faithfully recover optical phases from single-shot intensity measurements. To resolve the phase ambiguity without typical approximations~\cite{Kadlecek2001,Turner2005,Sobol2014,Smits2020,Altuntas2021}, we apply 
a general iterative algorithm~\cite{Sobol2014} to reconstruct coherent forward scattering from atomic samples illuminated by a precisely pre-characterized wavefront~\cite{Gerchberg1972,Ivanov1992,Latychevskaia2019}. By comparing the reconstructed coherent scattering with the unperturbed probe, the complex optical response is obtained for the simultaneous inference of the atomic density distribution and the velocity field. We also demonstrate single-shot spectroscopic imaging across EIT resonance with photon-shot-noise limited accuracy for the phase angle of the transmitted slow light, even in presence of large shot-to-shot atom number fluctuations. Based on Gabor's holographic microscopy (GHM)~\cite{Gabor1971} with intrinsic phase stability, the inline method can be conveniently set up to assist a wide range of cold atom experiments to access phase space information with in situ and minimally destructive measurements. Our imaging technique also paves a practical pathway toward precise spectroscopic imaging in presence of density fluctuations generic to most ultra-cold atomic samples~\cite{Gomez2006, Lu2013, Marti2018, Li2020Wu}.

   \begin{figure*}[htbp]
     \centering
     \includegraphics[width=.8 \textwidth]{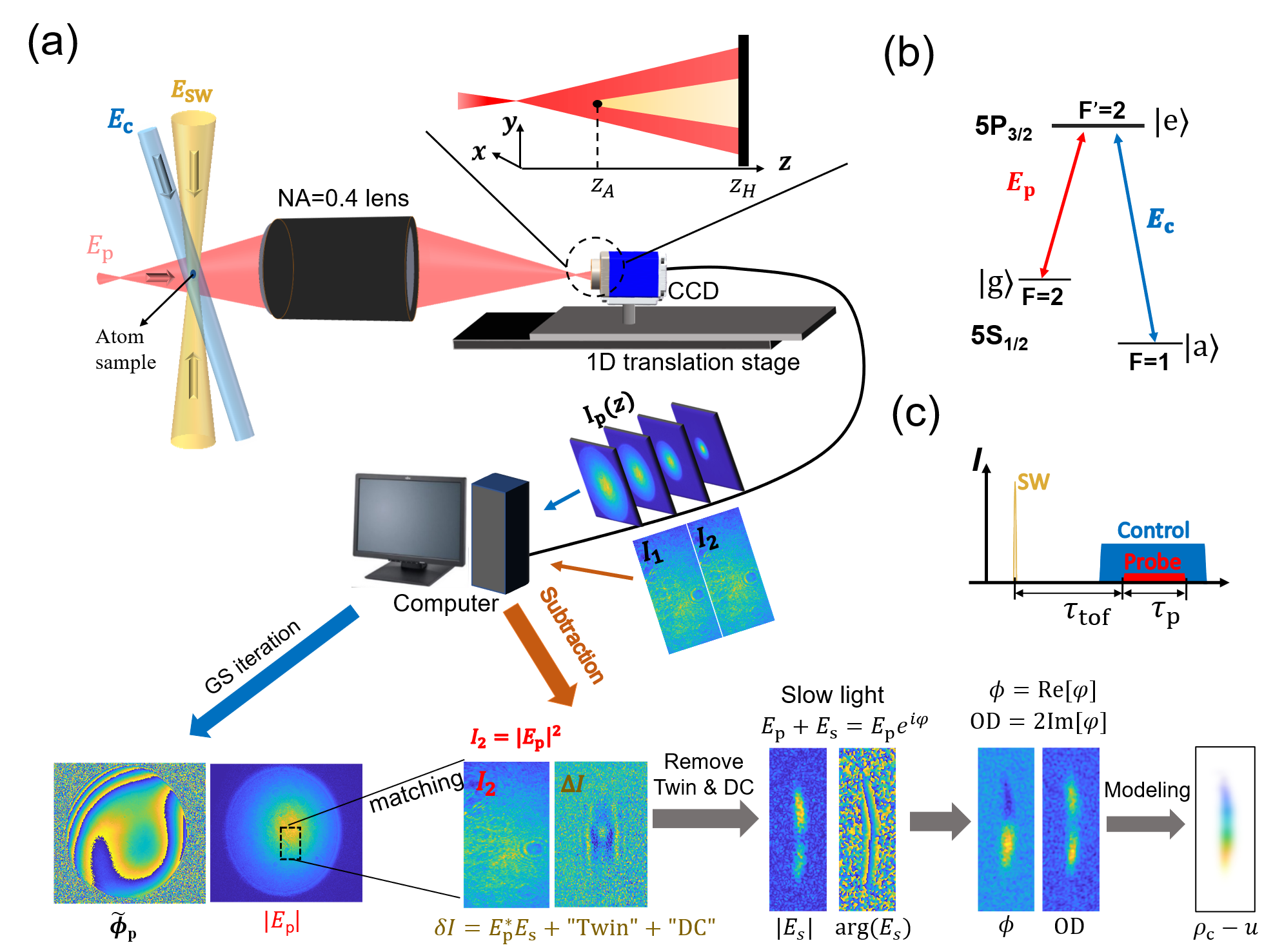}
     \caption{Holographic sensing of slow light wavefront. (a) Experimental setup and holographic reconstruction procedure: A sample of $^{87}$Rb atoms centered at $z=z_{\rm A}$ is illuminated by a defocused probe light $E_{\rm p}$ collected by an $NA=0.4$ imaging optics. By mounting the CCD camera on a linear stage, the probe beam $E_{\rm p}$ can be pre-characterized by multi-plane intensity measurements. 
     The EIT condition (with the level diagram in (b)) is established by first illuminating the atoms with the $E_{\rm c}$ control beam and then the $E_{\rm p}$ probe beam. A controllable velocity distribution is achieved by subjecting atoms to $E_{\rm SW}$ standing wave pulses along $y$. The camera data $I_{1,2}$ with and without atoms are matched to the precisely pre-characterized $E_{\rm p}=|E_{\rm p}|e^{i\phi_{\rm p}}$, including aberrations and speckles, to retrieve the atomic coherent scattering $E_{\rm s}$. The slow light wavefront and complex transmission $e^{i\varphi}=1+E_{\rm s}/E_{\rm p}$ can then be derived at $z_{\rm A}$ to retrieve the distribution of optical depth ${\rm OD}=2{\rm Im}[\varphi]$ and phase shift $\phi={\rm Re}[\varphi]$, and to infer the atomic column density $\rho_{\rm c}$ and velocity field $u$ simultaneously. See caption of Fig.~\ref{fig:accStandingWave} for the color coding of the $\rho_{\rm c}-u$ plot.
     For clarity of display the $\tilde{\phi}_{\rm p}\equiv \phi_{\rm p}-\phi_{\rm sp}$ plot gives the phase difference between $\phi_{\rm p}$ and the optical phase of a best-matched spherical wave $\phi_{\rm sp}$. 
     (c) The timing sequence. \label{fig:EITsetup} }
   \end{figure*}

 \section{Transmission of dragged slow light}
 
We consider the experimental setup illustrated in Fig.~\ref{fig:EITsetup}a. A cold $^{87}$Rb atomic sample  centered at $z=z_{\rm A}$ plane is illuminated by a $\lambda_p=780$~nm probe beam $E_{\rm p}$ with frequency $\omega\approx \omega_{eg}$ resonant to the $F=2 - F'=2$  hyperfine D2 transition. $E_{\rm p}$ also serves as a reference beam to interfere with the coherently scattered light from the atomic sample, $E_{\rm s}$, to be recorded on the $z=z_{\rm H}$ plane. The EIT condition is established for $F=2$ ground state atoms by illuminating them with a control beam $E_{\rm c}$ with frequency $\omega_c=\omega_{e a}$ resonant with the $F=1 - F'=2$ hyperfine transition. 

The transmitted wavefront of the weak probe light through the atomic medium can be expressed as $E_{\rm out}=E_{\rm p}+E_{\rm s}$. At the exit side of the atomic sample, $E_{\rm out}(z_{\rm A})$ is related to $E_{\rm p}(z_{\rm A})$ through a simple relation, $E_{\rm out}(z_{\rm A},\omega)=e^{i \varphi(\omega)}E_{\rm p}(z_{\rm A},\omega)$, for samples with moderate thickness $L$ in the Raman-Nath regime.
 Here and in the following we may omit the coordinate variables $x,y$ in the $E(z, \omega)$, $\varphi(\omega)$ expressions for conciseness. For a dilute gas, the complex phase $\varphi(\omega)$ is given by
   \begin{equation}
     \varphi(x,y,\omega)=\int_{z_{\rm A}-L}^{z_{\rm A}} \frac{\chi({\bf r},\omega)}{2}\frac{\omega}{c} \dd z, \label{eq:specPhase}
   \end{equation}
   which is divided into $\varphi(x,y,\omega)=\phi+i {\rm OD}/2$ to represent the phase shift $\phi(x,y)$ and optical depth ${\rm OD}(x,y)$. We consider the atomic phase space distribution $f({\bf r},{\bf v})$ and the associated spatial density $\rho({\bf r})=\int f({\bf r},{\bf v})\dd^3{\bf v}$. Recall that EIT is associated with coherent spin-wave excitation with wavevector ${\bf k}_{\rm eff}={\bf k}_{\rm p}-{\bf k}_{\rm c}$, with ${\bf k}_{\rm p}$, ${\bf k}_{\rm c}$ the wavevectors of the probe and control fields locally seen by the atoms respectively~\cite{Fleischhauer2001}. The Doppler-shift leads to an inhomogeneously broadened polarizability,
 $ \chi({\bf r},\omega)=\int \dd^3 {\bf v} f({\bf r},{\bf v})\alpha(\omega-{\bf k}_{\rm eff}\cdot {\bf v})$,  where $\alpha(\omega)$ is the atomic polarizability with the imaginary part $\alpha_{\rm i}(\omega_0)$ minimized at the 2-photon resonance frequency $\omega_0=\omega_c-\omega_{g a}$. Here $\omega_{g a}=2\pi\times6.835$~GHz is the hyperfine splitting between the $F=1,2$ ground states~\cite{Steck2003}. We consider $\tilde\omega_0=\omega_0 +{\bf k}_{\rm eff}\cdot {\bf v_0}$ to be a Doppler-shifted EIT resonant frequency and Taylor-expand $\alpha(\omega-{\bf k}_{\rm eff}\cdot {\bf v})$ to have
   \begin{equation}
     \varphi(x,y,\tilde\omega_0)\approx \varphi^{(0)}(x,y)+\delta \phi(x,y),\label{eq:specPhase2}
   \end{equation}
   with
   \begin{equation}
   \varphi^{(0)}(x,y)=\int_{z_{\rm A}-L}^{z_{\rm A}} \rho({\bf r}) \frac{\alpha(\omega_0)}{2} \frac{\omega_0}{c} \dd z,\label{eq:phaseMin}
   \end{equation}
   and
   \begin{equation}
   \delta \phi(x,y)=\int_{z_{\rm A}-L}^{z_{\rm A}} \rho({\bf r})\frac{\omega_0}{2}\frac{ d\alpha}{d \omega } \frac{{\bf k}_{\rm eff}\cdot \delta {\bf u}({\bf r})}{c} \dd z.\label{eq:phaseV}
   \end{equation}
 From the complex transmission $e^{i\varphi}=E_{\rm out}/E_{\rm p}$ one shall try to retrieve information on the velocity field 
   $\delta {\bf u}({\bf r})={\bf v}_0-{\bf u}({\bf r})$,  with
   \begin{equation}
   {\bf u}({\bf r})=\frac{1}{\rho({\bf r})}\int \dd^3 {\bf v}  f({\bf r},{\bf v}) {\bf v},\label{eq:u}
   \end{equation}
   relative to a pre-estimated ${\bf v}_0$ frame. While $\varphi_0$ vanishes at the 2-photon resonance for ideal EIT, practically $\varphi^{(0)}=i{\rm OD}^{(0)}/2+\phi^{(0)}$
   may arise with residual absorption and phase shifts due to various mechanisms including non-ideal EIT preparation and ground state relaxation~\cite{Novikova2012}. By setting the probe frequency $\omega$ near $\tilde \omega_0$, photon recoil heating during imaging can be managed to a suitable level for minimally destructive measurement of $\varphi(\omega)$ at within the EIT window. 


We now remark on the relation between Eq.~(\ref{eq:phaseV}) and the light-drag effect. In particular, the velocity-dependent phase shift $\delta \phi(x,y)$ is real, and is a manifestation of the light-drag effect that modifies the light phase velocity ${\bf v}_{\rm p}$, $\delta {\bf v}_{\rm p}\approx n_{\rm g} \delta {\bf u}({\bf r})$, in the highly dispersive medium with group index $n_{\rm g}({\bf r})\approx \rho({\bf r})\frac{\omega}{2}\frac{\dd\alpha}{\dd\omega}$~\cite{Safari2016,Kuan2016, Chen2020, Solomons2020}. If the velocity field $\delta {\bf u}({\bf r})$ is $z$-independent, then $\delta \phi(x,y) \approx {\bf k}_{\rm eff}\cdot \delta {\bf u}(x,y) \tau(x,y)$ with $\tau(x,y)=\int_{z_{\rm A}-L}^{z_{\rm A}} \dd z/v_{\rm g}({\bf r})$ the optical delay introduced by the atomic medium.  One may try to estimate ${\bf u} (x,y)$ along the ${\bf k}_{\rm eff}$ direction with the optical $\delta \phi(x,y)$ readouts and additional $\tau(x,y)$ measurements. However, in typical cold atom experiments the local atomic density $\rho({\bf r})$ is not precisely known. 
On the other hand, GHM demonstrated in this work is able to reconstruct $E_{\rm out}(z_{\rm A})$ and $E_{\rm p}(z_{\rm A})$ directly to spectroscopically determine the 2-photon Doppler shift ${\bf k}_{\rm eff}\cdot {\bf u}(x,y)$, so as to recover the velocity field ${\bf u}(x,y)=-\delta {\bf u}(x,y)+{\bf v}_0$ along ${\bf k}_{\rm eff}$. 
   

 \section{Holographic reconstruction}

Our holographic imaging method follows the philosophy of replacing standard approximations~\cite{Cuche1999,2004microelectromechanical, Bjrn2005Investigation,2008Phase,Greenbaum2012,DeHaan2020,Kadlecek2001,Turner2005,Sobol2014,Smits2020,Altuntas2021} with measurements whenever possible.  The measurement procedure to be detailed in ref.~\cite{suppInfo} is composed of four steps (Fig.~\ref{fig:EITsetup}a). First, before the holographic measurements, the probe wavefront $E_{\rm p}$ is precisely pre-characterized with multiple z-plane $I_{\rm p}(z)=|E_{\rm p}(z)|^2$ measurements. The 2D phase information of $E_{\rm p}(z_{\rm H})=\sqrt{I_{\rm p}(z_{\rm H})}e^{i\phi_{\rm p}(z_{\rm H})}$, $\phi_{\rm p}$, is recovered with an iterative Gerchberg-Saxton algorithm~\cite{Gerchberg1972,Ivanov1992}. Here, for the nearly spherical $E_{\rm p}$, the deviation ($\bar\phi_{\rm p}$ in Fig.~\ref{fig:EITsetup}) is mainly caused by aberration through the imaging system and sources of speckles in the imaging path. Next, $I_{1,2}$ are recorded with and without the atomic sample respectively. Numerical translation of $E_{\rm p}$ is applied to minimize the difference between $I_2$ and $|E_{\rm p}|^2$, thereby establishing the relation $I_1=|E_{\rm p}+E_{\rm s}|^2$ and $I_2=|E_{\rm p}|^2$ even in presence of slow alignment drifts. In the third step, $E_{\rm s}$ is retrieved from $\Delta I=I_1-I_2$ with an iterative algorithm~\cite{Sobol2014}, applicable to general probe beam wavefronts, to progressively removes both the twin and dc images~\cite{Jensen1997,Sobol2014,Sobol2014a} based on the  precise knowledge of $E_{\rm p}$ and known location of the atomic sample. Also in this step, the actual imaging resolution ${\rm NA}<{\rm NA}_0$ is flexibly chosen by digitally reducing the exposure area of the camera for the reduced hologram $\Delta I$. In the final step, we numerically propagate $E_{\rm p}$ and $E_{\rm s}$ to the $z_{\rm A}$ plane to retrieve the complex phase $\varphi(x,y,\omega)=-i {\rm log}(\frac{E_{\rm p}+E_{\rm s}}{E_{\rm p}})$. Regular ${\rm OD}(x,y)=2{\rm Im}[\varphi]$ and $\phi(x,y)={\rm Re}[\varphi]$, as well as ``dark ground" $|E_{\rm s}(x,y)|^2$ imaging data~\cite{Pappa2011} are simultaneously retrieved. In addition, we obtain the less often discussed phase angle $\beta(x,y)={\rm arg}[\varphi]$ and modulus $|\varphi(x,y)|$ images. 

In the GHM setup sketched in Fig.~\ref{fig:EITsetup}a, an ${\rm NA}_0$=0.4 imaging system~\cite{LI2018} relays both the probe light $E_{\rm p}$ and the scattered wavefront from the atomic sample $E_{\rm s}$ to the $z=z_{\rm A}$ plane.
The nearly spherical $E_{\rm p}$~\cite{Gabor1971} leads to a reduction in the probe intensity at the hologram plane $I_{\rm p}(z_{\rm H})=|E_{\rm p}(z_{\rm H})|^2$ relative to the intensity $I_{\rm p}(z_{\rm A})=|E_{\rm p}(z_{\rm A})|^2$ effectively seen by the atoms, a feature that has been utilized to enhance the detector dynamic range limited atomic sensing capacity~\cite{Sobol2014}. Here, 
to ensure plenty of camera counts $I_{1,2}$ within a short probe interval $\tau_{\rm p}$, we choose $z_{\rm A}=1.6$~mm and $z_{\rm H}=3.2$~mm with a moderate enhancement factor of $(z_H/z_A)^2=4$.



\section{Complex atomic spectroscopy}\label{sec:spec}
\begin{figure}[htbp]
  \centering
  \includegraphics[width=0.5 \textwidth]{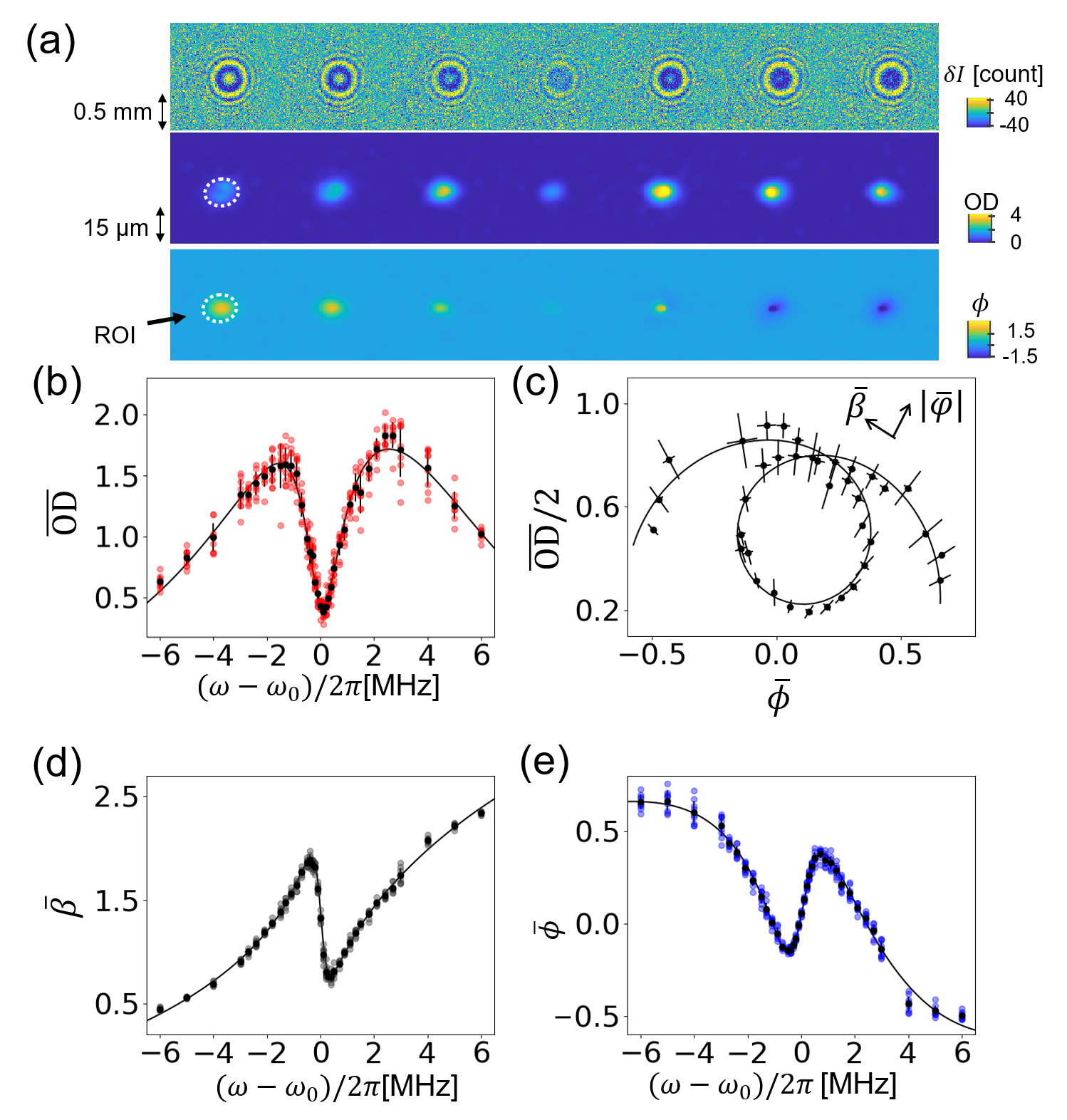}
  \caption{\label{fig:EITspectrum} 
   Spectroscopic imaging across the EIT resonance. (a) From top to bottom: Reduced holograms $\Delta I$ and the derived ${\rm OD}$ and $\phi$ images
  with probe detunings $\omega-\omega_0$ from $-6$~MHz to $6$~MHz with $\sim$2~MHz interval. An elliptical ROI with $\mathcal{A}_{\rm ROI}\approx $ 130~$\mu$m$^2$ is marked with white dotted lines. The images are from a sub-set of single-shot data presented in Figs.~(b-e). Figs.~(b)(d)(e) are ROI-averaged spectra for optical depth ($\overline{\mathrm{OD}}$), phase angle ($\overline{\beta}$) and phase shift ($\overline{\phi}$) respectively. Single-shot data are plotted with semi-transparent symbols to illustrate the statistical noise level, while the mean value for $N_m=9$ are plotted with dark dots. The data is further plotted as a phasor diagram in (c), with 2D errorbars giving the standard deviations along the $|\bar\varphi|$ and $\beta$ directions. 
   The solid curves are fit to a phenomenological EIT model~\cite{suppInfo}.}
\end{figure}

 We first demonstrate shot-noise limited spectroscopic imaging of stationary atomic sample in presence of fluctuating atom number. Here the probe beam is set with an intensity parameter $s\approx 2.2$ ($s=I/I_{\rm s}$ with saturation intensity $I_{\rm s}=1.67~{\rm mW/cm}^2$) at the location of the atoms. Holograms are recorded with an exposure time of $\tau_{\rm p}=20~\mu$s. The atomic sample is composed of $N_A\approx 1\times 10^4$ atoms, nearly spherical with a diameter $L\approx 13~\mu$m and temperature $T\approx 15~\mu$K. The control beam has an intensity of $\sim 7$~mW/cm$^2$ to maintain an EIT window width $\delta \omega_{\rm EIT}$ of 1~MHz. Holograms are recorded in repeated experiments with the probe frequency $\omega$ scanning across the 2-photon resonance $\omega_0$. Typical reduced holograms $\Delta I=I_1-I_2$, with $\bar{I}_{1}\approx 300$ in terms of camera counts on an $A_{\rm p}=6.45\times6.45~\mu$m$^2$ pixel area are given in Fig.~\ref{fig:EITspectrum}a (top) (Intensity of the probe for recording $I_2$ is increased by a factor of 25 to reduce the shot-noise contribution.).  In addition to the expected rings of interference signals, one clearly see the granular photon shot noise background, which is at a 20-counts root-mean-square level.  
 Following the described procedure, we match $I_2$ to $|E_{\rm p}|^2$, and use $E_{\rm p}$ to recover $E_{\rm s}$ from single holograms to retrieve the absorption ${\rm OD}(x,y)$ and phase shift $\phi(x,y)$.  By averaging the simultaneously reconstructed ${\rm OD}$ and $\phi$ images as in Fig.~\ref{fig:EITspectrum}a over a region of interest (ROI) that covers most of atomic signals, we obtain EIT curves for average absorption $\overline {\rm OD}$ (Fig.~\ref{fig:EITspectrum}b), phase shift $\bar\phi$ (Fig.~\ref{fig:EITspectrum}e), phase angle $\bar\beta={\rm arg}(\bar\varphi)$ (Fig.~\ref{fig:EITspectrum}d) and phasor diagram (Fig.~\ref{fig:EITspectrum}c) of $\{\bar{\phi}, \overline{\rm OD}/2\}$ plots~\cite{Cronin2004}. From the phase shift data in Fig.~\ref{fig:EITspectrum}(e), in particular, we can estimate an average group index of $n_g\sim 3\times 10^6$ and a reduced group velocity of light $v_{\rm g}$ at the 100~m/s level~\cite{VestergaardHau1999} across the sample within the $\sim$1~MHZ EIT window.

Unless working with confined single atoms~\cite{Marti2018}, a standard challenge in cold atom laser spectroscopy~\cite{Gomez2006, Lu2013, Li2020Wu} is associated with fluctuating atom number and coupling strengths during the measurements. Here, the spectroscopy is recorded by repeatedly preparing the cold atomic samples. The shot-to-shot atom number fluctuation translates into a spreading of data points in both $\overline{\rm OD}$ (Fig.~\ref{fig:EITspectrum}b) and $\overline{\phi}$(Fig.~\ref{fig:EITspectrum}e) spectroscopy. Looking into individual $\overline{\rm OD}$ or $\overline{\phi}$ data, it is impossible to isolate such fluctuation from frequency-dependent atomic response of interest. To efficiently suppress the noise, one may have to resort to normalizing the atom number with double measurements as carefully as possible~\cite{Li2020Wu}. Here, a unique strength of the complex spectroscopy is to unveil the correlation of such fluctuation, as demonstrated in the phasor diagram Fig.~\ref{fig:EITspectrum}(c) where the EIT resonance leads to a circle of complex atomic response. Unlike the photon shot noise, the spreading of the spectroscopic data in the Fig.~\ref{fig:EITspectrum}(c) phasor diagram is highly annisotropic since the shot-to-shot atom number fluctuation merely shifts the data points along the $|\overline{\varphi}|$ direction.


\begin{figure*}[htbp]
  \centering
  \includegraphics[width=0.85 \textwidth]{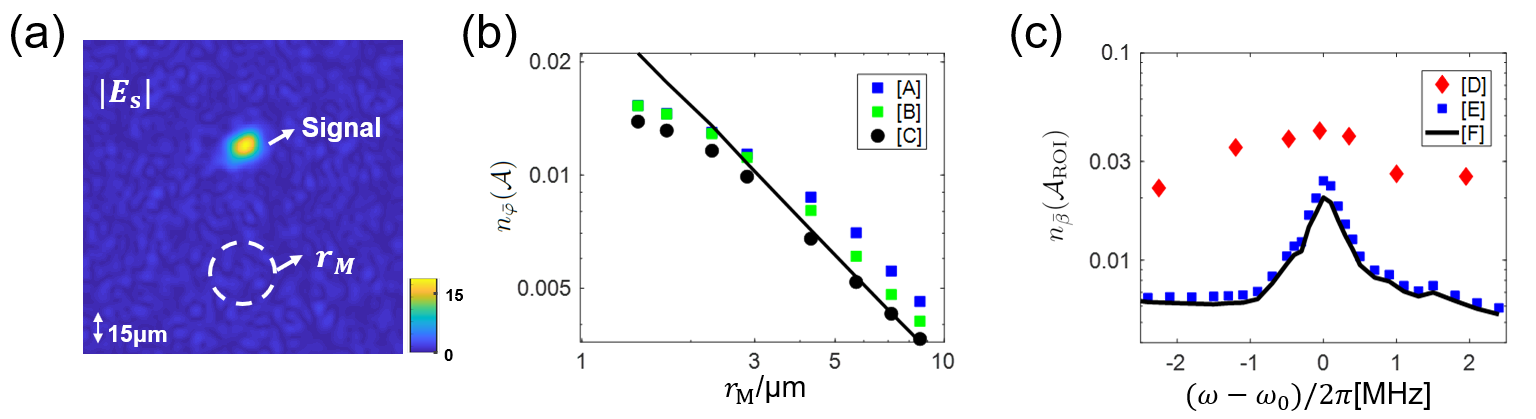}
  \caption{\label{fig:noiseF} Statistical analysis of the image noise. (a) Illustrated with a $|E_{\rm s}|$ image, the noise level $n_{\overline{\rm OD}}(\mathcal{A})$ and $n_{\overline{\phi}}(\mathcal{A})$ are evaluated from 100 complex-$\varphi$ images as $n_{\overline{\varphi}}(\mathcal{A})
  =\sqrt{n_{\overline{\rm OD}}^2/4+n_{\bar\phi}^2}$, within area $\mathcal{A}$ excluding the atom sample. (b): Dependence of $n_{\bar\varphi}(\mathcal{A})$ on radius $r_M$: Blue square ([A]) and green square ([B]) symbols are based on experimental $\varphi$-data measurements in presence and absence of the atom sample respectively. Black disks ([C]) are based on simulated shot-noise limited data. The black line gives $n=1/\sqrt{N_\mathrm{p}(\mathcal{A})}$, with $N_\mathrm{p}(\mathcal{A})$ to be the total number of photon counts through area $\mathcal{A}$ detected by the camera. Notice here for $r_M$ smaller than half the NA=0.13-limited $\delta x=6~\mu$m, the imaging noise are numerically smoothed down, leading to deviation of the data points from the power law, or even below the shot noise limit. (c): Noise of the phase angle $\overline{\beta}$ for atomic polarizability $\alpha$, estimated by $\varphi$-data within $\rm ROI$. 
  The red diamond symbol ([D]) gives root-mean-square value of the  shot-to-shot $\overline{\beta}$ fluctuation  estimated from groups of 36  measurements. Blue squares ([E]) give $n_{\bar\beta}(\mathcal{A}_{\rm ROI}) =n_{\overline{\varphi}}/\sqrt{2}|\bar\varphi| $ estimated according to the [A] method in Fig.~(b). The black line ([F]) gives $n_{\bar\beta}(\mathcal{A}_{\rm ROI})=\xi/\sqrt{N_{\rm s}(\mathcal{A}_{\rm ROI})}$, with $\xi=|(e^{-i\overline{\varphi}}-1)/\overline{\varphi}|$ to be an optical depth dependent correction factor~\cite{suppInfo}.}
\end{figure*}

We now focus on the phase angle $\overline{\beta}={\rm arg}(\overline{\varphi})$, which can be resolved by the complex spectroscopy from single shot measurements. For the dilute and optically thin samples, $\bar\beta$ is directly decided by the phase angle of the atomic polarizability $\alpha(\omega)$ (Eq.~(\ref{eq:specPhase})) and is insensitive to the atom number fluctuation. Indeed, the shot-to-shot noise level for $\overline{\beta}$ in Fig.~\ref{fig:EITspectrum}(d) within the EIT window is close to being photon-shot-noise limited. The shot-noise-limited performance is demonstrated in Fig.~\ref{fig:noiseF}. Here, with $|\varphi|\ll1$ and therefore $\varphi\approx -i E_{\rm s}/E_{\rm p}$ at the EIT resonance, the phase angle measurement is equivalent to interferometrically estimating the optical phase of the forward scattering $E_{\rm s}$ relative to the reference field $E_{\rm p}$~\cite{ScullyBook} through an aperture defined by the ROI. It is known that the holographic microscopy is a ``spatial heterodying" measurement~\cite{Kadlecek2001}. Therefore, the phase angle measurement here has a resolution bounded by $n_{\bar\beta}> 1/\sqrt{N_{\rm s}(\mathcal{A}_{\rm ROI})}$, with $N_{\rm s}(\mathcal{A}_{\rm ROI})$ to be the amount of elastically scattered photons by the atoms through ROI to be detected by the camera in term of interference with $E_{\rm p}$. More generally, for optically thick samples we have $n_{\bar\beta}>\xi/\sqrt{N_{\rm s}(\mathcal{A}_{\rm ROI})}$, with $\xi=|(e^{-i\overline{\varphi}}-1)/\overline{\varphi}|$ to be an optical depth dependent correction factor~\cite{suppInfo}. We verify the shot-noise-limited phase angle resolution by directly evaluating the total number of scattered photons from the reconstructed $E_{\rm s}$ field at $z=z_{\rm A}$ as $N_{\rm s}(\mathcal{A}_{\rm ROI})=\sum_{\rm ROI}|E_{\rm s}|^2$ after proper normalization (Fig.~\ref{fig:noiseF}c, the line plot)~\cite{suppInfo}. For $\omega\approx\omega_0$ with optimal EIT transparency we typically find $N_{\rm s}(\mathcal{A}_{\rm ROI}) \approx 2000$, predicting shot-noise-limited $n_{\overline{\beta}} \approx 0.02$. The shot-to-shot $\bar\beta$ fluctuation as displayed by the diamond symbols in Fig.~\ref{fig:noiseF}(c)  is twice as large at the EIT resonance. The lightly noisier $\bar\beta$ is likely caused by power fluctuation of EIT control field ${\bf E}_c$ in repeated measurements which affects $\alpha(\omega)$ itself~\cite{suppInfo}. 
With increased light scattering when $\omega-\omega_0$ is off the EIT resonance, the $\bar\beta$ fluctuation deviates further from the $\xi/\sqrt{N_{\rm s}(\mathcal{A}_{\rm ROI})}$ prediction, due to the technical noise.
 
The precision of the phase-angle measurement is further characterized by estimating $n_{\bar\beta}=n_{\overline{\varphi}}/\sqrt{2}|\bar\varphi|$ (Fig.~\ref{fig:noiseF}c  the square symbols) in an atomic sample independent manner using individually reconstructed complex data of $\varphi=\phi+i{\rm OD}/2$. The method assumes the noise level in each $\rm OD$, $\phi$ images is approximately constant at the vicinity of the sample location ROI. As illustrated in  Fig.~\ref{fig:noiseF}(a), with the probe photon shot noise dominating the holography noise channels and with $|z_H-z_A|\gg \mathcal{A}/\lambda$ for the shot noise to digitally propagate~\cite{Sobol2014}, the standard deviation $\{n_{\bar\phi}(\mathcal{A}), n_{\overline{\rm OD}}(\mathcal{A})/2\}$ evaluated as root-mean-square values from repeated measurements is found to be quite isotropic and can thus be simultaneously characterized by the noise level to the complex phase $n_{\overline{\varphi}}\equiv\sqrt{n^2_{\bar\phi}+n^2_{\overline{\rm OD}}/4}\approx \sqrt{2} n_{\overline{\phi}}$. From Figs.~\ref{fig:noiseF}(a)(b) we see $n_{\overline{\varphi}}$ itself is close to the probe photon shot limit $1/\sqrt{N_{\rm p}}$~\cite{Sobol2014}, decided by the amount of probe photons $N_{\rm p}(\mathcal{A})$ through the imaging area $\mathcal{A}$ detected by the camera. Here, with $n_{\overline{\varphi}}(\mathcal{A}) \approx 1.3/\sqrt{N_{\rm p}(\mathcal{A})}$, the slightly larger $n_{\overline{\varphi}}(\mathcal{A})$ is associated with additional technical noises, and may also be related to imperfect digital holographic processing such as those during $I_{1,2}$ subtraction~\cite{suppInfo}. With the shot-noise-limited $n_{\bar\varphi}$, the phase angle noise level in Fig.~\ref{fig:noiseF}c (the square plot) evaluated as $n_{\bar\beta}(\mathcal{A}_{\rm ROI}) =n_{\overline{\varphi}}(\mathcal{A}_{\rm ROI})/\sqrt{2}|\bar\varphi(\mathcal{A}_{\rm ROI})|$ (with modulus $|\bar\varphi(\mathcal{A}_{\rm ROI})|=\sqrt{\overline{\phi}^2+\overline{\rm OD}^2/4}$ evaluated within ROI) is close to the $1/\sqrt{N_{\rm s}(\mathcal{A}_{\rm ROI})}$ shot noise limit, as expected.

 \section{Velocity field sensing}
 
We create atomic samples with spatially dependent velocity field ${\bf u}({\bf r})$ by subjecting the dipole trapped atoms with standing wave diffraction, followed by a $\tau_{\rm tof}$ expansion along $y$ in a quasi-2D trap with $x-z$ confinements (Fig.~\ref{fig:EITsetup}). The standing wave is formed along $y$ by $\lambda_{\rm a}=795~$nm counter-propagating D1 light (Fig.~\ref{fig:EITsetup}a) blue-detuned from the $F=1-F'=2$ hyperfine resonance by 1~GHz. The pulse is powerful ($I_{\rm a}=10{^3}$W/cm$^2$) and short ($\tau_{\rm a}=50~$ns) for precise modeling in this investigation. Microscopically, a periodic repulsive light shift potential is formed by the optical standing wave for the bidirectional acceleration of atoms along $y$.  Macroscopically, after the optical impact the phase space distribution of atomic sample along $y$ is characterized by $f(y,v_y)=\rho(y)g(v_y)$ with a velocity distribution $g(v)$ determined by the sinusoidal standing wave shift. The control field ${\bf k}_{\rm c}$ is aligned in the $y-z$ plane with a small angle $\iota=11.1^{\circ}$ relative to ${\bf e}_y$~\cite{He2020a}. The nearly free expansion leads to $f(y,v_y,t)=f(y-v_y t,v_y)$, resulting in a velocity field ${\bf u}(y,t)$ (Eq.~(\ref{eq:u})) approximately along $y$.

\begin{figure}[htbp]
  \centering
  \includegraphics[width=0.5 \textwidth]{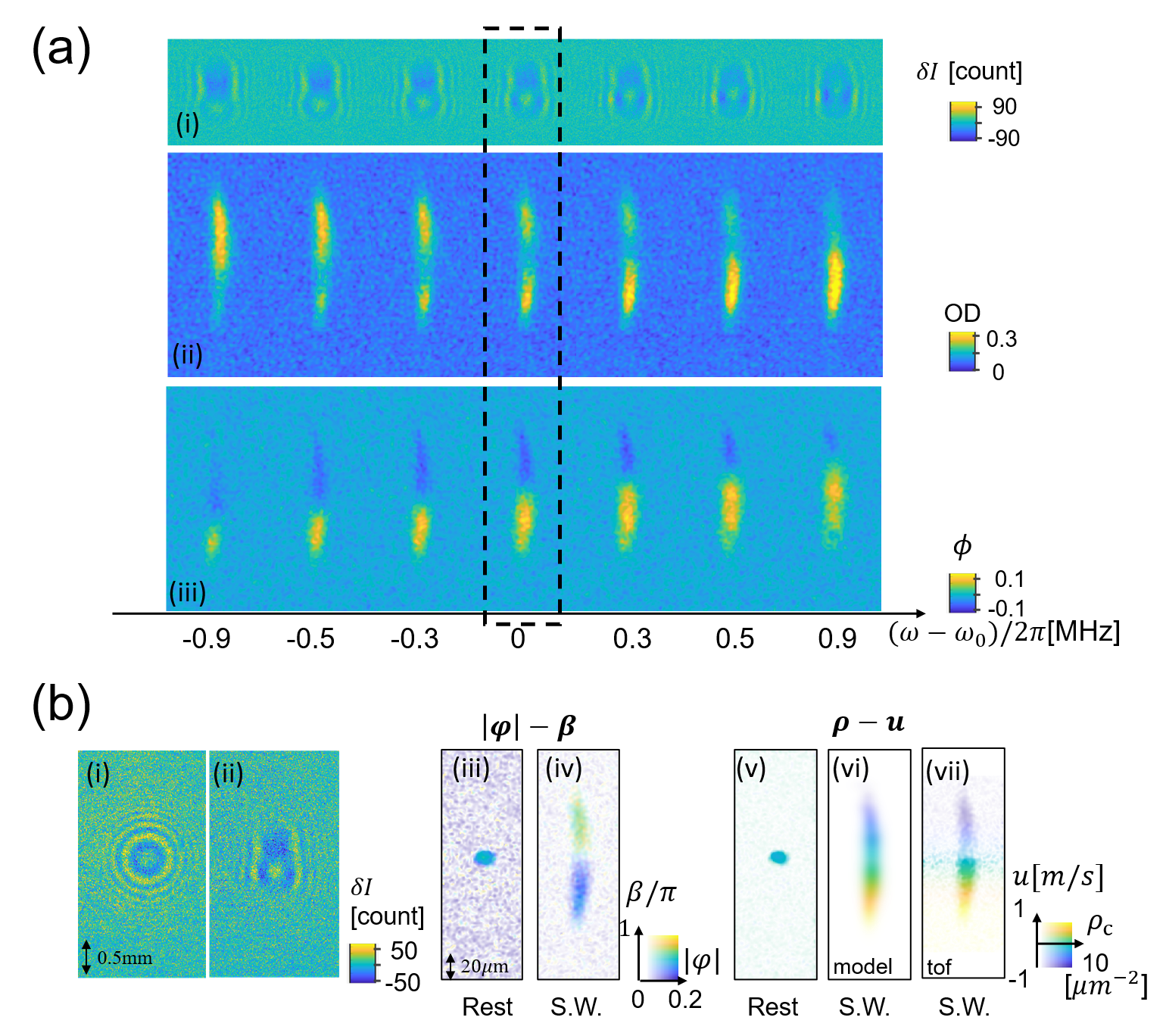}
  \caption{Holographic reconstruction of spatial-dependent light drag effect for velocity field sensing. 
  (a) From top to bottom: Reduced holograms $\Delta I$ and the derived ${\rm OD}$ and $\phi$ images  for the standing-wave-accelerated atomic samples. The holograms are taken near the EIT condition with the two-photon detuning $\delta$ scanned across the EIT resonance. Each shot-noise-limited hologram is averaged 80 times to enhance the signal to noise for the display. (b) Reduced holograms for atomic sample at rest (i) and expanding along $y$ (ii), recorded with zero 2-photon detuning. The reconstructed complex-$\varphi$ data are displayed with color-domain images in  (iii)(iv). From the $\varphi$-data, smooth $\rho_c-u$ distributions are inferred in (v,vi). Direct $\rho_c-u$ retrieval assuming $u_y(y)$ following the one-source expansion is given in Fig.~(vii)~\cite{suppInfo}. 
  Here the velocity is coded in color and atomic column density is coded in brightness.  \label{fig:accStandingWave}}
\end{figure}

We probe the expanding atomic sample after $\tau_{\rm tof}=20~\mu$s for a $\tau_{\rm p}=10~\mu$s exposure time. The exposure is short enough that most atoms hardly move a distance beyond the numerically-reduced NA$\approx 0.13$-limited imaging resolution $\delta x\approx 6~\mu$m~\cite{suppInfo}. The probe field $E_{\rm p}$ transmitted through the expanding sample is holographically recorded by the camera. 
We retrieve the phase shift holographically as the real part of the complex phase $\varphi$ (Fig.~\ref{fig:accStandingWave}(a,iii)), together with the absorption (Fig.~\ref{fig:accStandingWave}(a,ii)).  For a certain probe frequency $\omega$, a particular velocity class of atoms satisfying $\omega-\omega_0={\bf k}_{\rm eff}\cdot {\bf v}_0$ falls into the EIT window with enhanced transmission. Around the EIT window, the phase shift $\phi$ always increases linearly with $y$ for all the probe frequency $\omega$, suggesting the same $u(y)-y$ dependence as expected from the point-source expansion. 
With a reduced peak atomic density $\rho$ for the expanding cloud at a $5\times 10^{11}/{\rm cm}^3$ level, the group velocity of the slow light is increased to $v_{\rm g}\approx 1000$~m/s, and it takes $\tau=L/v_{\rm g} \approx 20$~ns for the probe pulse to trespass the atomic sample at the EIT center. The optical phase of the probe wavefront locally follows the atomic motion, leading to $\delta \phi(x,y)=k_u u(x,y)\tau$ phase shift in the transmitted wavefront for 
${\bf v}_0=0$ in particular. The ${\bf u}$-component of ${\bf k}_{\rm eff}$ assuming ${\bf u}=u {\bf e}_y$ is $k_u=k_{\rm c}{\rm cos}(\iota)+\epsilon k_{\rm p}$, with $\epsilon={\bf k}_{\rm p}\cdot {\bf e}_y/k_{\rm p}$ numerically evaluated from the $E_{\rm p}$ wavefront at the atomic location. For $|u| <0.3$~m/s within the linear dispersion regime, the maximum shift $\phi_{\rm M}$ is at a 50~mrad level, suggesting a dragged transverse wavefront displacement by $(\delta y)_{\rm M}=\phi_{\rm M}/2\pi \lambda_{\rm p}\approx 5$~nm during the slow light travel. However, without a prior knowledge of the atomic density distribution, precise characterization of the atomic motion cannot be realized with the $\phi(x,y)$ image alone. 

Here, with the imaging data expanded from real to complex numbers by holography (See the $|\varphi|-\beta$ color domain plot in Fig.~\ref{fig:accStandingWave}(b,iii-iv).), we have the opportunity to estimate atomic density $\rho$ and the velocity component $u$ simultaneously from single shot measurements. In particular, for the thin-lens regime of light scattering by the dilute sample, the atomic column density $\rho_{\rm c}(x,y)$ and velocity field $u(y)$ can be inferred from the $\varphi(x,y)$ data by minimizing $\mathcal{L}=\sum_{x,y} |2 \varphi(x,y)-k_{\rm p} \rho_{\rm c} \alpha(\omega-k_u u)|^2 $ ~\cite{suppInfo}. The atom polarizability $\alpha(\omega)$ is obtained beforehand by fitting a simple EIT model to the spectroscopic measurements as those in Fig.~\ref{fig:EITspectrum}. Typical inferred $\rho_{\rm c}-u$ fields are given in Figs.~\ref{fig:accStandingWave}(v,vi) probed at $\omega=\omega_0$ for the stationary and expanding atomic clouds respectively. For comparison, we plot velocity field $u=u_y(y)$ in Fig.~\ref{fig:accStandingWave}(vii)  estimated directly as $u_y=y/(\tau_{\rm tof}+1/2\tau_{\rm p})$ following the point source expansion~\cite{suppInfo}. Notice a curved $\rho_{\rm c}$ distribution suggests that the pulsed standing wave light shift induces certain ``slosh'' motion of atoms along $x$ during the expansion in the dipole trap. We expect similar motion along $z$ which is close to ${\bf k}_{\rm p}$ to be EIT-sensitive. The small discrepancies between the Fig.~\ref{fig:accStandingWave}(b,vi-vii) velocity field estimations could be due to the additional $z$-motion sensed by the EIT spectroscopy not to be recorded by the displacement along $y$ in the same images.

\section{Summary and outlook}  
   

In this work, we have demonstrated a technique to precisely reconstruct the full wavefront of slow light transmitted through cold atomic samples with single-shot intensity measurements. We have shown that as in other coherent imaging schemes~\cite{Lye2003,Sobol2014}, the quality of the reconstructed spectroscopic imaging data is close to that limited by the shot noise of the elastically scattered photons from atoms, and can thus be  minimally destructive for velocity field sensing. The key advantage of the holographic imaging method is the ability to retrieve absorption and phase shift simultaneously. By expanding the imaging data from real to complex numbers, the mixed information on the velocity field and density distributions can be disentangled and separately inferred. 

Comparing with previous works~\cite{Cuche1999,2004microelectromechanical, Bjrn2005Investigation,2008Phase,Greenbaum2012,Sobol2014, DeHaan2020}, our method is a simplest application of Gabor's original holography~\cite{Gabor1971}, but with various pre-assumptions of unknowns removed for better accuracy. Future development of this holographic spectroscopic imaging approach, particularly toward imaging through optics with larger numerical aperture~\cite{Bakr2009,Sherson2010b}, may offer a platform that combines high spatial resolution~\cite{Bakr2009,Sherson2010b,Boll2016,Yang2020} with interferometric precision for spectroscopic imaging~\cite{Marti2018} and velocity sensing~\cite{Moler1992, Matsko2003, Su2010, Carey2019} with ultracold atoms.


   
   
\section*{Funding Information}
 National Key Research Program of China (2017YFA0304204, 2016YFA0302000); 
 National Science Foundation of China (11574053). Shanghai Scientific Research Program (15ZR1403200). 
 
 
 
 \section*{Appendix}

 \subsection{Holographic retrieval of probe transmission and atomic response}

A key technique in this work is to reconstruct the transmitted wavefront $E_{\rm out}=E_{\rm p}+E_{\rm s}$ to be compared with $E_{\rm p}$ so as to simultaneously obtain the atomic absorption ${\rm OD}(x,y)$ and phase shift $\phi(x,y)$ from single-shot holographic measurements. 
The holographic technique is composed of four steps as following.

First, before the atomic imaging measurements, we pre-characterize the wavefront of the probe field, $E_{\rm p}$, through multi-plane intensity measurements followed by a multi-plane Gerchberg-Saxton (G-S) algorithm to reconstruct the phase information~\cite{Gerchberg1972,Ivanov1992}. 

Next, for the holographic imaging, defocused images $I_1=|E_{\rm p}+E_{\rm s}|^2$ and $I_2=|E_{\rm p}|^2$ are recorded at a ``holography'' plane $z_{\rm H}$, with and without the atomic sample respectively, as outlined in the main text (Fig. 1). Taking into account ambient noise that affect both $I_1$ and $I_2$, in this step a numerical procedure is applied to optimally match $I_2$ to  the  pre-characterized $E_{\rm p}$ and to derive the reduced hologram $\Delta I=I_1-I_2$.

Thirdly, $E_{\rm s}$ is numerically reconstructed from the reduced hologram $\Delta I=I_1-I_2$ with an iterative algorithm to remove the twin and dc noises~\cite{Sobol2014a}.

Finally, $E_{\rm s}$ and $E_{\rm p}$ are numerically  propagated to atomic plane $z=z_{\rm A}$ to evaluate the complex phase $\varphi(x,y)=\phi+i {\rm OD}/2=-i {\rm log}(\frac{E_{\rm p}(z_{\rm A})+E_{\rm s}(z_{\rm A})}{E_{\rm p}(z_{\rm A})})$. 

As in the main text, we may omit the $x,y$ (and $k_x,k_y$) coordinate variables in the notions of wavefronts and intensities if ambiguity would not be introduced.

To propagate a wavefront $E(x,y,z)$ from $z=z_{\rm A}$ to $z=z_{\rm B}$, we use the angular spectrum method with the following notations:
\begin{equation}
\begin{array}{l}
E(z_{\rm B})=\hat U(z_{\rm B}-z_{\rm A})E(z_{\rm A}),{\rm with}\\
\hat{U}(L)={\hat{F}}^{-1}e^{i  \sqrt{k^2-k_x^2-k_y^2}L}\hat{F}
\end{array}\label{eq:asm}
\end{equation}
Here $k=2\pi/\lambda$ is the wavenumber of the light field, and $\hat{F}$ represent the 2D Fourier transform~\cite{kim2010principles}, $i.e.$, $\hat F E(x,y,z)=E(k_x,k_y,z)$ and $\hat F^{-1} E(k_x,k_y,z)=E(x,y,z)$.

In the following we detail the four-step holographic reconstruction procedure.

\subsubsection{Pre-characterization of $E_{\rm p}$}

\begin{figure}[htbp]
    \centering
    \includegraphics[width=\linewidth]{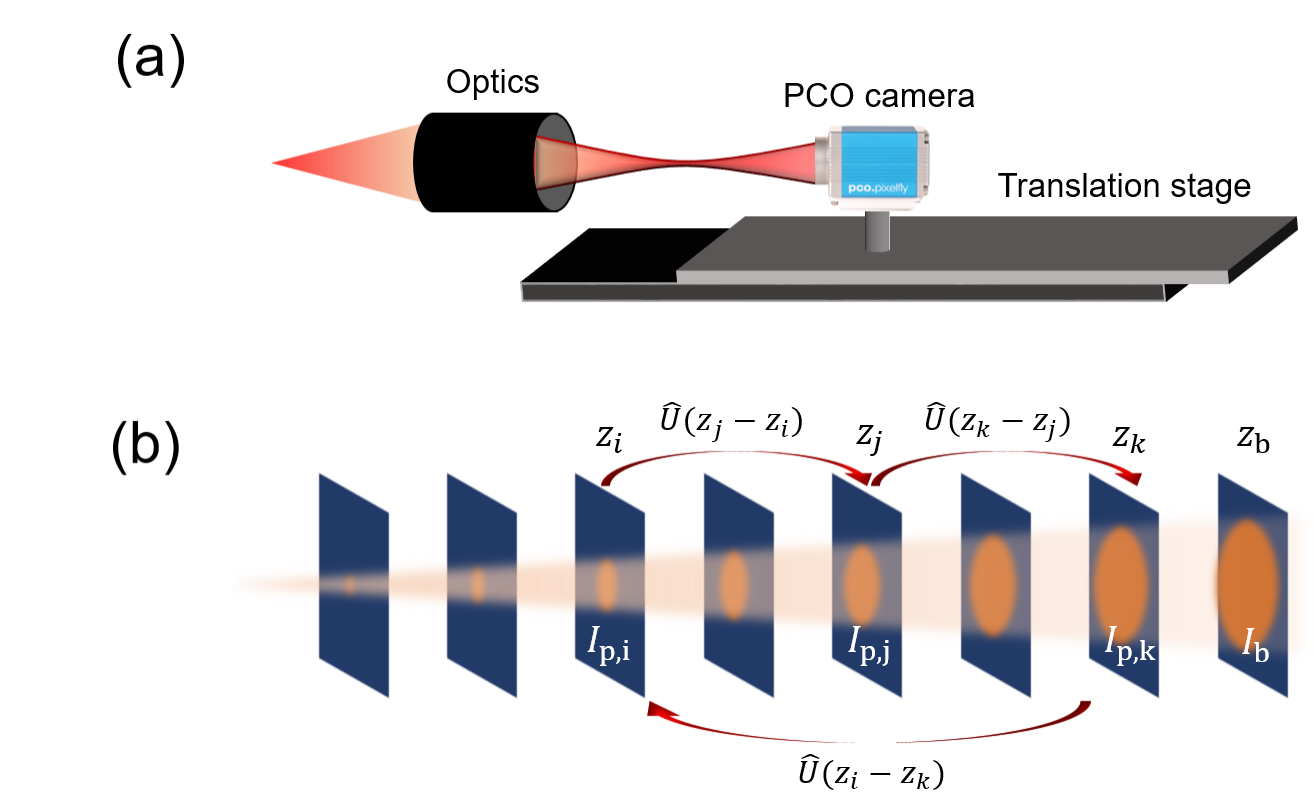}
    \caption{Schematic setup for the measurement (a) and numerical reconstruction (b) of the probe light wavefront $E_{\rm p}$. The intensity profile $I_{\rm p}(z_j)=|E_{\rm p}(z_j)|^2$ is recorded at multiple $z-$planes with a digital camera mounted on a translation stage. The flow diagram in (b)  describes a single 3-plane G-S iteration for the complex $E_{\rm p}$ reconstruction. The full reconstruction is composed of multiple 3-plane G-S iterations with $\{I_{\rm p,j}\}$ data randomly chosen from the pre-recorded data-set. }
    \label{FigS-GS1} 
\end{figure}

As in Fig.~\ref{FigS-GS1}(a), we use a CCD camera mounted on a stepper-motor-driven linear stage to sample the probe field intensity distribution $I_{\rm p}(x,y,z)==|E_{\rm p}(x,y,z)|^2$ in 3D, by translating the camera along $z$ and record a set of 2D intensity profiles $I_{\rm p,j}=I_{\rm p}(z_j)$. Here the $x-y$ plane is defined by the camera sensor chip. The direction normal to the plane is thus referred to as the $z-$direction, which is adjusted to be parallel to the direction of translation to within $\sim$10~mrad angular precision. At each $z_j$-plane, the intensity $I_j$ is adjusted with an acoustic-optical modulator (AOM) to ensure sufficient counts on the camera without saturation. The distance $z_j$ is digitally recorded, which is subjected to numerical adjustments as detailed in the following. The imaging data is referred to as a $\{I_{\rm p}\}$-set. 

The 3D intensity sampling is subjected to systematic noises such as multiple reflections involving the camera sensor chip itself. To recover $E_{\rm p}$ from the imperfect $\{I_{\rm p}\}$ measurement, it is important to sample many $z_j-$ planes to suppress the position-dependent noise so as to determine $E_{\rm p}$ through the average intensity field.  To achieve the purpose without demanding computer memory, we apply a multi-plane G-S algorithm that samples the $\{I_{\rm p}\}$-set in small groups, detailed as following. It should be noted that although we take a strongly focused probe beam (NA=0.15 after all the imaging optics) as the example in this work, the method is applicable to reconstructing general form of probe beam wavefronts.

\begin{enumerate}
\item[(1)] Split pixels: each image of the camera-recorded $I_{\rm p}$-set, as represented by a $M\times N$ 2D matrix, is split by a factor of $n_{\rm s}\times n_{\rm s}$ to enhance the spatial resolution $\Delta x=\Delta x^{(0)}/n_{\rm s}$. Here $\Delta x^{(0)}$ is the physical camera sensor resolution. For the CCD camera in this work, with $M=1240$, $N=1392$, and $\Delta x^{(0)}=6.45~\mu$m, we find $n_{\rm s}=4$ is large enough to support the complex $E_{\rm p}$ reconstruction at $NA=0.15$. We choose $n_{\rm s}=6$ with large enough margin in this work.

\item[(2)] Set a ``bank'' plane: We choose a ``bank'' plane $z_{\rm b}$ and the associated probe beam intensity profile $I_{\rm b}$ from the $\{I_{\rm p}\}$-set. The phase of the wavefront $E_{\rm b}$ is initialized as a spherical wave phase $\phi_{\rm p}^{\rm s w}=k\sqrt{x^2+y^2+z_{\rm b}^2}$. To do this, we have assumed $z=0$ for the $E_{\rm p}$ focal plane, which is estimated by additional camera images near $z=0$. An initial guess of $E_{\rm p}$ as $E_{\rm p}^{G S} = \sqrt{I_{\rm b}} e^{i\phi_{\rm p}^{\rm s w}}$ starts the G-S iteration in the next step. With $z_{\rm b}$ chosen as the ``bank'' plane, $I_{\rm b}$ is subsequently removed from the $\{I_{\rm p}\}$ data-set.

\item[(3)] Three images from the $\{I_{\rm p}\}$-set are randomly chosen with intensity distribution $I_{\rm p,i},I_{\rm p,j},I_{\rm p,k}$ recorded at $z_i,z_j,z_k$.  With the angular spectral method by Eq.~(\ref{eq:asm}), $E_{\rm p}^{\rm GS}$ is first propagated to $z_i$ as $E_{\rm p}^{\rm GS} \leftarrow \hat{U}(z_i-z_{\rm b}) E_{\rm p}^{\rm GS}$.

\item[(4)] The following G-S iteration among the $i,j,k$ planes are applied  for $n_{\rm A}$ times as following (Fig.~\ref{FigS-GS1}b): Cyclically propagate $E_{\rm p}^{\rm GS}$ from $z_i$ to $z_j$, $z_k $ and back to $z_i$. In each step and before the next numerical propagation,  the amplitude of $E_{\rm p}^{\rm GS}$ is replaced with the camera-recorded $\sqrt{I_{\rm p}}$ while keeping the phase information. We usually set $n_{\rm A}=30$.

\item[(5)] 
With the $z_{i,j,k}$ values initially estimated from the linear stage readouts, the $z_{i,j,k}$ optimization procedures are sparsely inserted into the G-S iteration in step 4. For example, when $E_{\rm p}^{\rm GS}$ is propagated from $z_i$ to $z_j$, the value $z_j$ is adjusted so as to optimize the inner product between the $|E_{\rm p}^{\rm GS}(z_j)|^2$ and $I_{\rm p, j}$ images. The corrections to $z_{i,j,k}$ are updated in the following iterations.

\item[(6)] Add to the ``bank'' wavefront: Propagate $E_{\rm p}^{\rm GS}$ back to the ``bank'' plane $z_{\rm b}$. Add $E_{\rm p}^{\rm GS}$ to the ``bank'' waveform $E_{\rm p}^{\rm Bank}$ with an exponential weight: 
$E_{\rm p}^{\rm Bank} \leftarrow \frac{1}{1+\eta} ( E_{\rm p}^{\rm Bank} + \eta E_{\rm p}^{\rm GS})$. Here $\eta$ is a weight factor that determine the convergence speed and stability. After the bank-addition procedure, the G-S wavefront for the iteration is updated with the best estimated value at the ``bank'' location: $E_{\rm p}^{\rm GS} \leftarrow E_{\rm p}^{\rm Bank}$.

\item[(7)] Repeat the (3)-(6) loop $n_{\rm B}$ times until the reconstruction similarity $s=\langle \Delta I_{\rm t}|\delta |E_{\rm p}^{\rm GS}(z_{\rm t})|^2\rangle$, an ``inner product'' between the reconstructed $|E_{\rm p}^{\rm GS}(z_{\rm t})|^2$ and $I_{\rm t}$ profiles converge. Here $I_{\rm t}$ the intensity distribution at a target plane far away from the planes for the $\{I_{\rm p}\}$ set. The reconstructed reference wavefront is finally obtained at $z=z_{\rm b}$ as $E_{\rm p} = E_{\rm p}^{\rm GS}$, and is now free to propagate to any other $z$-plane.

\end{enumerate}

Here, with $\Delta I_{\rm A}=I_{\rm A}-\bar I_{\rm A}$ and $\Delta I_{\rm B}=I_{\rm B}-\bar I_{\rm B}$ the similarity is defined as
\begin{equation}
    \langle \Delta I_{\rm A}|\Delta I_{\rm B}\rangle=\frac{\overline{\Delta I_{\rm A} \Delta I_{\rm B}}}{\sqrt{\overline{\Delta I_{\rm A}^2}}\sqrt{\overline{\Delta I_{\rm B}^2}}}.
\end{equation}

\begin{figure}[htbp]
    \centering
    \includegraphics[width=\linewidth]{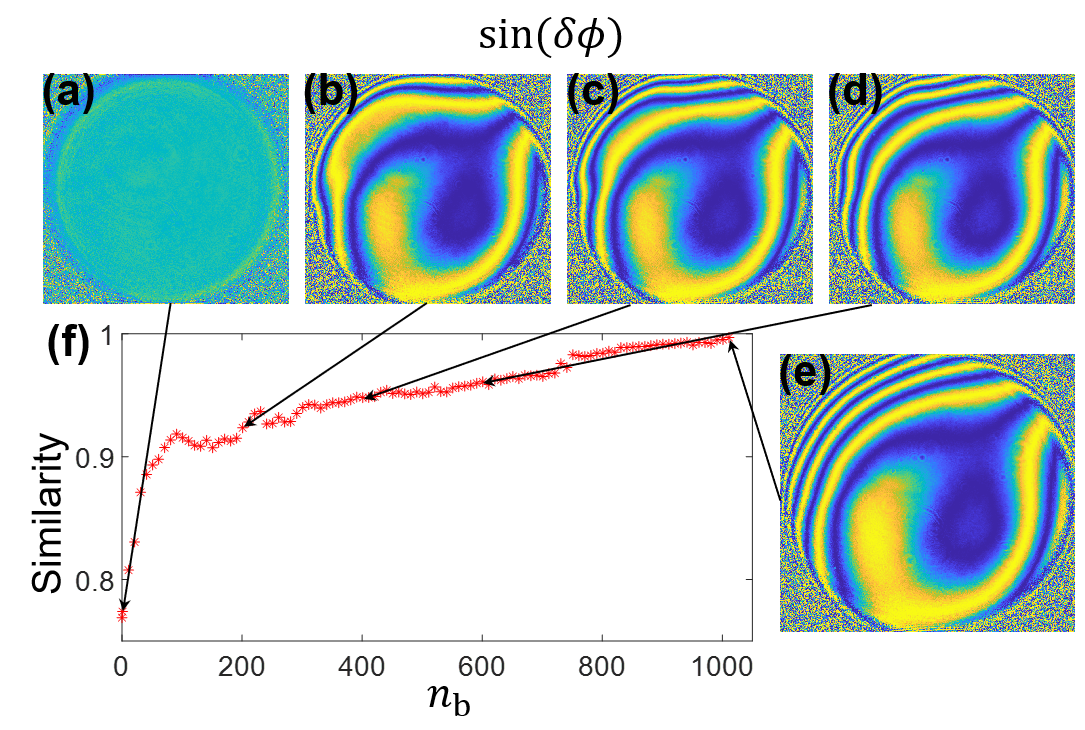}
    \caption{Evolution of the reconstruction fidelity $f$. Typical phase difference between iterated wavefront and the spherical wave $\delta\phi_{\rm p}=\phi_{\rm p}-\phi_{\rm p}^{\rm s w}$ (Fig. a-e) are also presented. The final wavefront }
    \label{FigS1-1} 
\end{figure}

In Fig.~\ref{FigS1-1} we plot the evolution of the similarity $s$ during a typical $E_{\rm p}$ reconstruction, together with the reconstructed wavefront phase $\phi_{\rm p}$ at typical steps of the G-S iteration. The G-S process typically take $n_{\rm b}=1000$ steps toward full convergence. The final intensity similarity $s\approx 0.99$ suggests a similar level of fidelity for the reconstructed wavefront $E_{\rm p}=\sqrt{I_{\rm p}}e^{i\phi_{\rm p}}$. The slow iteration is a result of low-efficiency correction to the large scale wavefront aberrations. In ref.~\cite{futurework} an optimization procedure that substantially speeds up the process, as well as a full characterization of the wavefront reconstruction fidelity will be discussed.

\subsubsection{Hologram recording, reduction and $E_{\rm p}$ matching\label{sec:Subtraction}}

In this step, holograms of atomic sample are recorded as $I_1$ and $I_2$, with and without the atomic sample respectively as outlined in the main text (Fig.~1). To suppress the impact of vibration noises during the $I_{1,2}$ recording, we minimize the inter-frame time to be merely 10~ms just to allow the atomic sample to escape the probe area. We then optimally subtract away the ambient background recorded by additional background measurements, following a procedure detailed in ref.~\cite{Sobol2014}. The procedure allows us to obtain the reduced hologram $\Delta I=|E_{\rm p}+E_{\rm s}|^2-|E_{\rm p}|^2$ for the $E_{\rm s}$ retrieval in the next step. To account for drifts of optics alignments, We perform a 3D numerical shift of $E_{\rm p}$ to optimally match the pre-characterized $|E_{\rm p}(z_{\rm H})|^2$ with $I_2$.

\subsubsection{$E_{\rm s}$ retrieval}

We now divide the reduced hologram $\Delta I$ with the optimally-matched $E_{\rm p}^*$ to obtain our initial guess of $E_{\rm s}$,
\begin{equation}
E_{\rm diff}=\frac{\Delta I}{{E_{\rm p}}^*}=E_{\rm s}+\frac{{E_{\rm s}}^*E_{\rm p}}{{E_{\rm p}}^*}+\frac{|E_{\rm s}|^2}{{E_{\rm p}}^*}
\label{Eq-2-2}
\end{equation}


\begin{figure}[htbp]
    \centering
    \includegraphics[width=\linewidth]{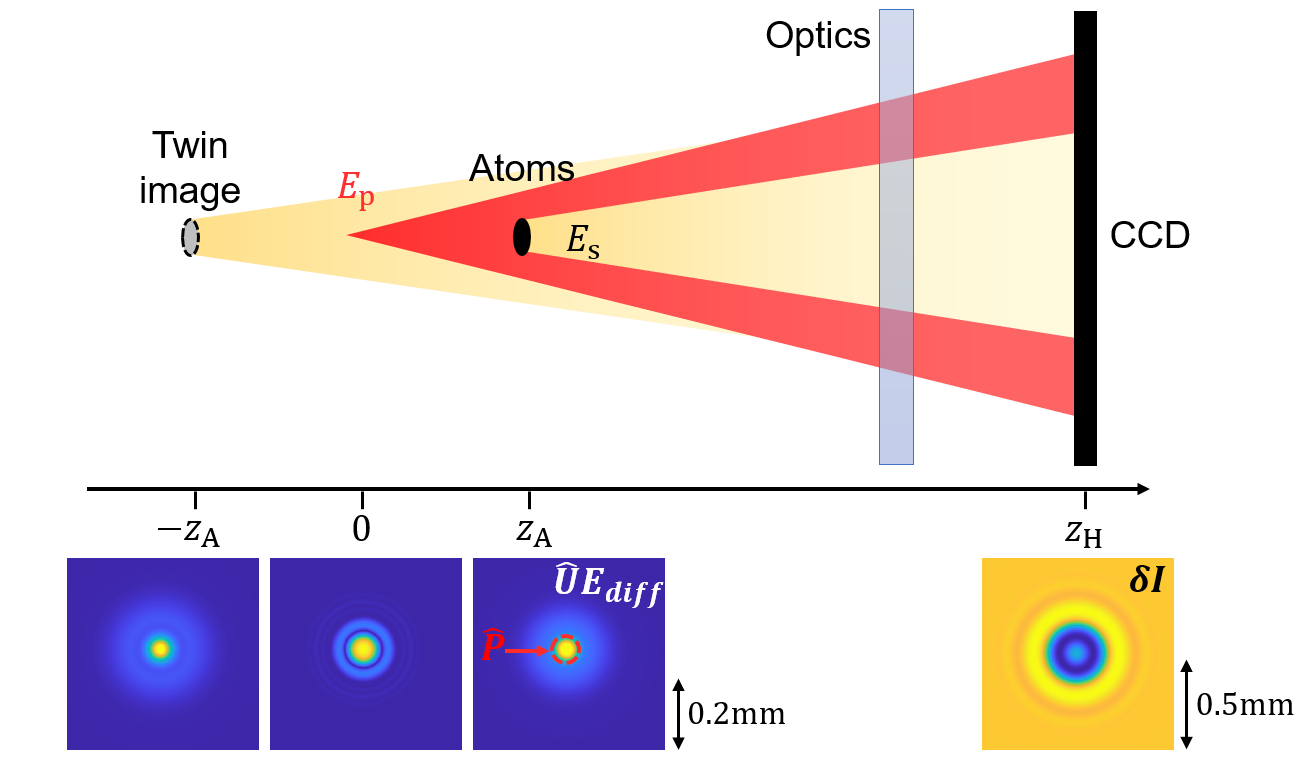}
    \caption{Schematic illustration of the $E_{\rm s}$, twin image, DC term. Here $E_{\rm s}$ is focused at atomic plane $z=z_{\rm A}$. DC and twin images are focused approximately at $z=0$ and $z=-z_{\rm A}$ respectively. To remove the twin and dc noise, a $\hat{P}$ operation can be introduced by picking up the $E_{\rm s}$ signal with concrete support.}
    \label{FigS2-1} 
\end{figure}
As illustrated in Fig.~\ref{FigS2-1}, $E_{\rm diff}$ contains information of $E_{\rm s}$ that properly focuses at the atomic sample plane $z=z_{\rm A}$. In addition, the twin-image term $E_{\rm s}^*E_{\rm p}/E_{\rm p}^*$ focuses approximately at $z=-z_{\rm A}$, while the dc noise $|E_{\rm s}|^2/E_{\rm p}^*$ focuses at $z=0$. Due to the inline geometry, these additional noise terms overlap with $E_{\rm s}$ and there is no simple method to isolate $E_{\rm s}$ without a prior knowledge. Here we employ minimal amount of a prior knowledge on the known location of the atomic sample, and use an iterative algorithm~\cite{Sobol2014} to remove the twin and dc noises and to faithfully retrieve $E_{\rm s}$ from $E_{\rm diff}$, as following:

First, we define an aperture operator $\hat P$, which sets $\hat P E(z_{\rm A})=E(z_{\rm A})$ inside the aperture and $\hat P E(z_{\rm A})=0$ outside the aperture (Fig.~\ref{FigS2-1}). With the concrete support of the $E_{\rm s}$ signal by the a prior location information, $\hat P E_{\rm s}(z_{\rm A})=E_{\rm s}(z_{\rm A})$,  $\hat P E_{\rm diff}(z_{\rm A})$ picks up all the $E_{\rm s}$ but removes part of the twin and dc noises. To iteratively improve the isolation of the $E_{\rm s}$, we take advantage of the symmetry between $E_{\rm s}$ and the twin-image terms to introduce a complex conjugation operator $\hat C$ at $z=z_{\rm H}$ plane as~\cite{Sobol2014}, 

\begin{equation}
\hat{C}E(z_{\rm H}) = \frac{E^*(z_{\rm H})E_{\rm p}(z_{\rm H})}{E_{\rm p}^*(z_{\rm H})}.
\label{Eq-2-4}
\end{equation}

A iteration procedure then follows
\begin{equation}
E_{\rm s}^{(n+1)} = \hat{U}\hat{C}(E_{\rm diff}-\hat{U}^{-1}\hat{P}E_{\rm s}^{(n)})
\label{Eq-2-3}
\end{equation}
to remove the twin and dc noise.

The initial value of of the $E_{\rm s}^{(n)}$ is set as $E_{\rm s}^{(n=0)} = \hat{U}E_{\rm diff}$. As discussed in ref.~\cite{Sobol2014,Sobol2014a}, $E_{\rm s}^{(n)}$ converges to $E_{\rm s}$ with a speed determined by the fraction of light field energy removed by $\hat P$ at each step.  Unlike the work in ref~\cite{Sobol2014} where a complex $E_{\rm p}$ composed of two point sources are introduced, here we simply use a nearly spherical $E_{\rm p}$. Fast convergence of the iteration is guaranteed by a small atomic sample with size $\sigma^2\ll \lambda z_{\rm A}$ to support a relatively small $\hat P$ $E_{\rm s}$ aperture. On the other hand, to image large atomic sample with $\sigma^2\sim \lambda z_{\rm A}$ we should increase the $E_{\rm p}$ complexity accordingly as those in ref.~\cite{Sobol2014}.

Finally, after the convergence of iteration by Eq.~(\ref{Eq-2-3}), the removal of the dc noise is achieved by updating the iteration relation as~\cite{Sobol2014}

\begin{equation}
E_{\rm s}^{(n+1)} = \hat{U}\hat{C}(E_{\rm diff}-\hat{U}^{-1}\hat{P}E_{\rm s}^{(n)} - \frac{|\hat{U}^{-1}\hat{P}E_{\rm s}^{(n)}|^2}{{E_{\rm p}}^*}).
\label{Eq-2-5}
\end{equation}

\begin{figure}[htbp]
    \centering
    \includegraphics[width=1\linewidth]{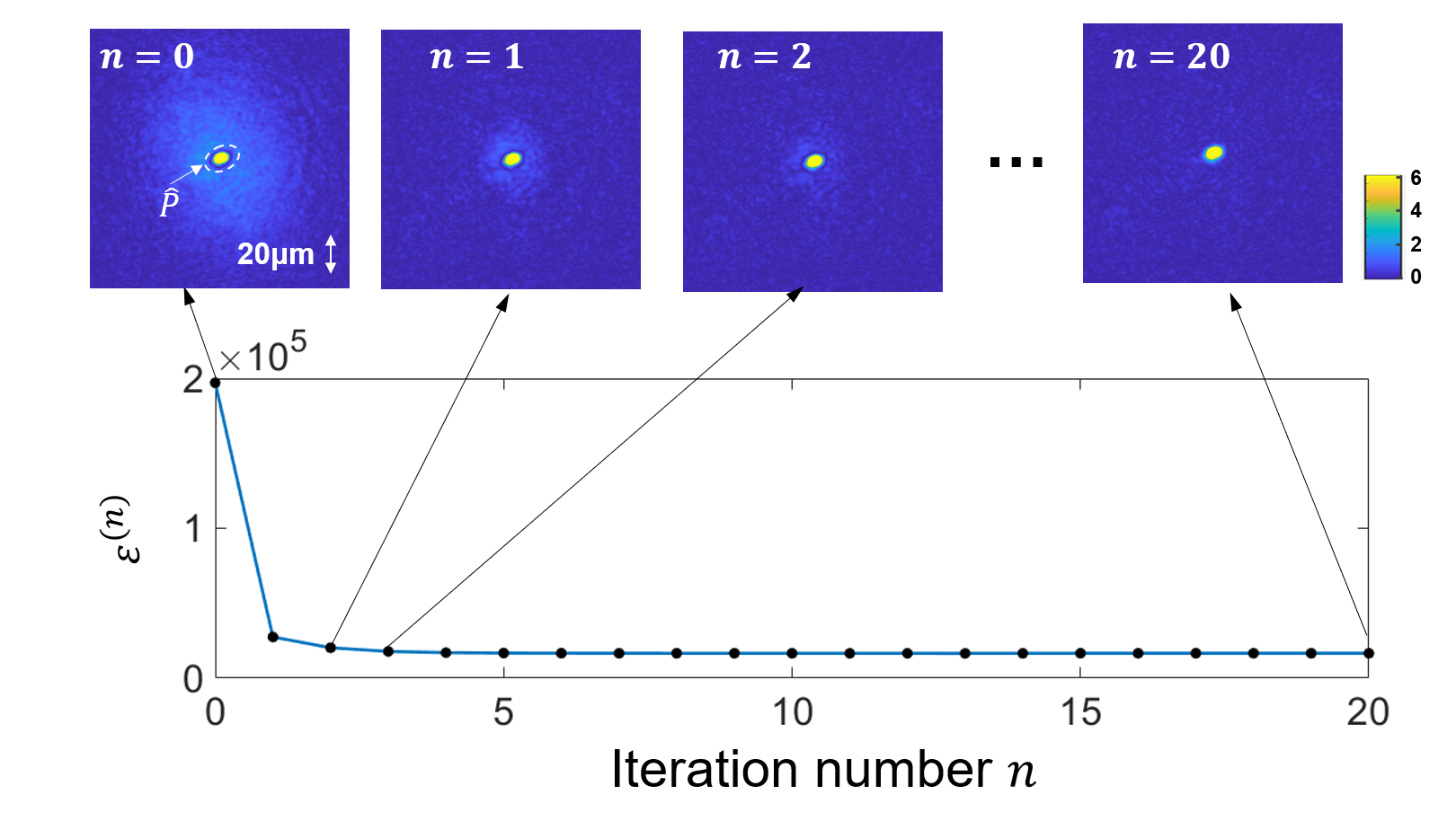}
    \caption{Convergence of the residual energy     
    $\varepsilon^{(n)}$ during the twin-DC removal iteration. Here $\varepsilon^{(n)}$ is summed over the displayed imaging area on the top, with the $\hat P$ area excluded.}
    \label{FigS2-2} 
\end{figure}
We characterize the convergence of iteration with a residual energy, $\varepsilon^{(n)}=\sum |(1-\hat P)E_{\rm s}^{(n)}|^2$ summing over an area substantially larger than the sample size at $z_{\rm A}$. Typical convergence of the residual is shown in Fig.~\ref{FigS2-2}.
Notably, the phase-recovery algorithm here does not assume any special form of reference wavefront $E_{\rm p}$, nor para-axial approximations that limit the spatial frequency of the wavefront. The method is therefore naturally compatible with the full wavefront reconstruction process based on precise $E_{\rm p}$ characterization.

\subsubsection{${\rm OD}$ and $\phi$ retrieval}

In this last step, we propagate the reconstructed $E_{\rm p}$ and $E_{\rm s}$ to the plane of atomic sample, $z=z_{\rm A}$, to obtain $\rm OD$ and $\phi$ data as the imaginary and real part of complex phase $\varphi(x,y)=-i {\rm log}(\frac{E_{\rm p}+E_{\rm s}}{E_{\rm p}})$. Here, with full knowledge of $E_{\rm p}$ and $E_{\rm s}$, various coherent imaging-processing steps can be taken to suppress imaging aberration and speckle noises. For example, to reduce the impact of speckle noises in $E_{\rm p}$ which are introduced by the imaging system itself (not seen by the atoms), we first propagate $E_{\rm p}$ to $z=0$ to limit its value within the spread of $E_{\rm s}$, before propagating the filtered $E_{\rm p}$ back to the $z=z_{\rm A}$ plane for the atomic response retrieval. It is worth noting these speckle noises should equally affect signals in regular absorption images where a similar correction is difficult to make.

\subsection{Details on the experimental setup and data analysis}


\subsubsection{Fitting the complex EIT spectrum}\label{sec:EIT}
\begin{figure}[htbp]
    \centering
    \includegraphics[width=0.5\linewidth]{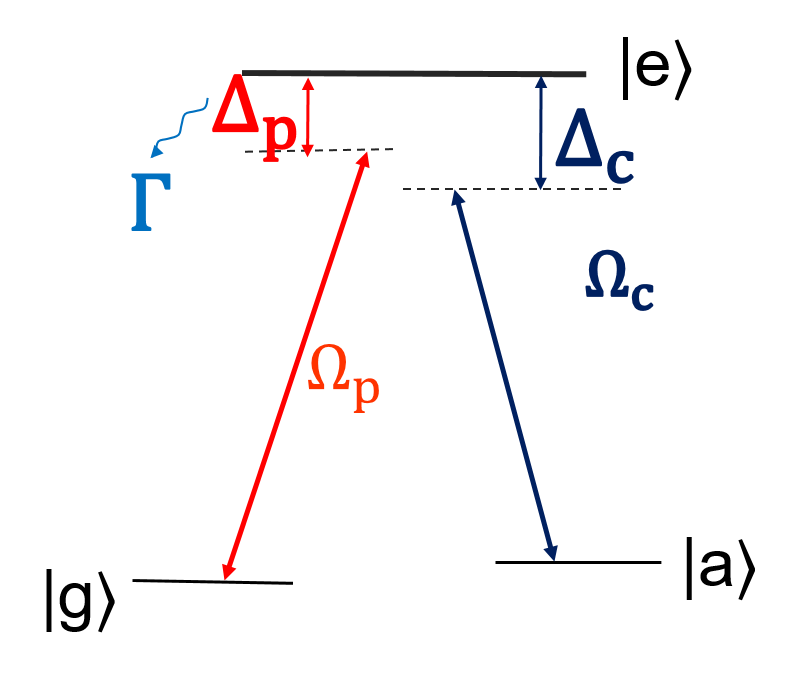}
    \caption{Schematic of a 3-level system system to obtain the phenomenological EIT response.}
    \label{fig:3level} 
\end{figure}

The complex spectrum obtained from the holographic measurements can be fit  to a phenomenological EIT model to assist the data analysis. The fit process is detailed in this section. Here, instead of a full analysis of EIT for the multi-level atom in presence of residual magnetic field,  we use a simple 3-level model (Fig.~\ref{fig:3level}) and perform a linear analysis of the atomic response to $E_{\rm p}$, leading to the standard expression of the atomic polarizability,

\begin{equation}
    \alpha(\omega_\mathrm{p},\omega_\mathrm{c})=\alpha_0+\frac{C}{\frac{|\Omega_\mathrm{c}|^2}{4(\Delta_\mathrm{p}-\Delta_\mathrm{c}-i\kappa/2)}-(\Delta_{\rm p}+i\Gamma/2) }.
  \label{eq:alpha-probe-2}
  \end{equation}
Here the parameter $C$  determines the $|g\rangle \leftrightarrow |e\rangle$ effective transition strength. The $\Delta_\mathrm{p}=\omega_\mathrm{p}-\omega_\mathrm{eg}$ and $\Delta_\mathrm{c}=\omega_\mathrm{c}-\omega_\mathrm{ea}$ are the probe and control laser detunings respectively. $\kappa$ is a phenomenological damping constant to account for relaxation between the ground state $|g\rangle $ and $|a\rangle$. A non-zero $\alpha_0$ is expected from real atomic response due to non-ideal optical pumping, additional magnetic coupling, as well as off-resonant coupling of $E_\mathrm{p}$ to additional hyperfine transitions.

For the dilute and optically thin atomic sample in this work, the  complex phase shift $\varphi(\omega_\mathrm{p},\omega_\mathrm{c})$ obtained from the holography is expected to be proportional to $ \alpha(\omega_\mathrm{p},\omega_\mathrm{c})$. We thus apply  Eq.~(\ref{eq:alpha-probe-2}) to fit the experimental complex phase from the $\mathrm{OD}$ and $\phi$ data in Figs. 2, 3, 4 of the main text for the case of stationary atoms. The free parameters 
$\alpha_0, C, |\Omega_\mathrm{c}|^2, \kappa, \Gamma, \Delta_\mathrm{c}$ for the Fig. 2 fit are given by: $\alpha_0=-0.026+0.38i$, $C=8.7$, $\Delta_\mathrm{c}/(2\pi)=-1.1$MHz, $\Gamma=13.7/(2\pi)$~MHz, $|\Omega_\mathrm{c}|^2/(2\pi)^2=11.7$~MHz$^2$, $\kappa/(2\pi)=-0.79$~MHz.

\subsubsection{Photon shot noise limit to the $\overline{\beta}$ measurement}


In the main text we considered optically thin samples which induce complex phase shifts $\varphi\ll1$, so that a measurement of phase angle $\beta={\rm arg}(\varphi)$ is equivalent to measuring the relative phase between $E_{\rm s}$ and $E_{\rm p}$, leading to $n_{\bar \beta}>1/\sqrt{N_{\rm s}}$ shot noise limit through an imaging aperture $\mathcal{A}$ in the atomic sample plane. 

Here we derive the more general bound $n_{\bar \beta}>\xi/\sqrt{N_{\rm s}}$ as in the caption of Fig.~3 in the main text. In particular, from the definition of the complex phase $\varphi=-i {\rm log}(1+E_{\rm s}/E_{\rm p})$, we have $\delta \overline{\varphi}=\delta \overline{E_{\rm s}}/(\overline{E_{\rm s}}+\overline{E_{\rm p}})$, with $\delta \overline{\varphi}$, $\delta \overline{E_{\rm s}}$ evaluated as the rms value of the complex number within an imaging area as those in Fig.~3a of the main text. When the imaging noise is dominated by the probe photon shot noise, we have $n_{\overline{\beta}}=|\frac{\delta\overline{\varphi}}{\sqrt{2}\overline{\varphi}}|$, and further $|\frac{\delta \overline{E_{\rm s}}}{\sqrt{2} \overline{E_{\rm s}}}|>1/\sqrt{N_{\rm s}}$. Following $\frac{\delta \varphi}{\varphi}=\frac{\delta \overline{E_{\rm s}}}{\overline{E_{\rm s}}}\frac{1-e^{-i\overline{\varphi}}}{\overline{\varphi}}$ we arrive at $n_{\overline{\beta}}>\xi/\sqrt{N_{\rm s}}$ with a optical-depth dependent factor $\xi=|\frac{1-e^{-i\overline{\varphi}}}{\overline{\varphi}}|<1$.

\subsubsection{Inference of $\rho_c-u$ distribution}

This section describes the procedure to infer atomic density $\rho_c$ and velocity field $u$ from the holographically reconstructed complex $\varphi$ data such as those for Fig.~4 in the main text. 

For the model based $\rho_{\rm c}-u$ inference leading to Fig.~4(vi), the numerical method parametrizes $\rho_c$ and $u$ fields to efficiently minimize the cost function $\mathcal{L}=\sum_{x,y}|2\varphi(x,y)-k_{\rm p}\rho_c (x,y) \alpha(\omega-k_u u(x,y))|^2$ where $\alpha(\omega)$ is obtained by fitting the EIT-data as described in Sec.~\ref{sec:EIT}.  Generally, assumptions are tailored into the parameters for specific type of sample characterizations. Here, taking advantage of a prior knowledge that the atomic density distribution $\rho_{\rm c}$ is localized and smooth, we decompose $\rho_{\rm c}$ into $N_{\rm G}$ Gaussian packets as $\rho_{\rm c}=\sum_{j=1}^{N_{\rm G}} C_j e^{-(x-X_j)^2/A_j^2-(y-Y_j)^2/B_j^2}$ parametrized by parameters $\{C_j, X_j, Y_j, A_j, B_j\}$. On the other hand, the velocity field $u(x,y)$ follows a polynomial expansion with $u(x,y)=\sum_{m,n=0}^{M,N} c_{m,n}x^m y^n$ parametrized by $\{c_{m,n}\}$. With in mind the limited signal/noise in single $\varphi(x,y)$ data, we restrict the parameters to $m=0$ and $n=0,1,2$ during the numerical optimization as following. 

The optimization procedure starts with $N_{\rm G}=1$ with quasi-Newton methods provided by Matlab (fminunc). After the $\mathcal{L}$ minimization converge, $N_{\rm G}$ is increased by one, with the new Gaussian packet inheriting the optimal parameters of the last Gaussian, except for a randomly shifted $X_{j}, Y_{j}$ combined with halved $A_{j}, B_{j}$ for a next round of nonlinear optimization. The process of ``splitting the last Guassian'' is continued up to $N_{\rm G}=7$ to obtain the model fit in Fig.~4(vi) in the main text, beyond which we find over-fitting of artifacts from apparent noises occurs. 

For the time-of-flight method leading to Fig.~4(vi), we assume that the atomic position recorded during the $\tau_{\rm tof}<t<\tau_{\rm tof}+t_{\rm p}$ camera exposure time are perfectly correlated to their velocity through the point-source expansion.  By setting $y=0$ for the unperturbed atomic position (Fig.~4(iii)), the velocity field is estimated as $u(y)=y/\tau_{\rm mean}$, with $\tau_{\rm mean}=\tau_{\rm tof}+1/2\tau_{\rm p}$ to be the average free-flight time. We then derive $\rho_{\rm c}(x,y)=2|\varphi(x,y)|/|k_{\rm p}\alpha(\omega-k_u u(x,y))|$ from the $\varphi(x,y)$ data.
 
 \bibliography{sl}

\begin{thebibliography}{58}
\bibitem [{fut()}]{futurework}%
  \BibitemOpen
  \href@noop {} {\bibinfo  {journal} {Since the preparation of this work, we
  have made additional developments to improve the speed of the probe wavefront
  pre-characterization, on phase-recovery for imaging large atomic sample with
  a non-iterative method, and on holography-based aberration corrections. The
  results will be given in a future publication.}\ }\BibitemShut {NoStop}%
\end{thebibliography}

 \end{document}


\preprint{APS/123-QED}

\title{Supplemental Material for \\
Imaging moving atoms by holographically reconstructing the dragged slow light}

\author{Yuzhuo Wang$^1$}
\email{zhuodashi@163.com}

\author{Jian Zhao$^1$}
\email{Current address: KLA-Tencor Semiconductor Equipment Technology (Shanghai) Co., Ltd.}

\author{Yizun He$^1$}
\author{Xing Huang$^1$}
\author{Lingjing Ji$^1$}
\author{Yudi Ma$^1$}
\author{Liyang Qiu$^1$}
\author{James P. Sobol$^2$}
\author{Saijun Wu$^1$}
\email{saijunwu@fudan.edu.cn}

\affiliation{$^1$Department of Physics, State Key Laboratory of Surface Physics and Key
Laboratory of Micro and Nano Photonic Structures (Ministry of Education),
Fudan University, Shanghai 200433, China.\\
$^2$Keit Spectrometers, Rutherford Appleton Laboratory, Oxfordshire UK.}

\date{today}


\maketitle

\section{Holographic retrieval of probe transmission and atomic response}

A key technique in this work is to reconstruct the transmitted wavefront $E_{\rm out}=E_{\rm p}+E_{\rm s}$ to be compared with $E_{\rm p}$ so as to simultaneously obtain the atomic absorption ${\rm OD}(x,y)$ and phase shift $\phi(x,y)$ from single-shot holographic measurements. 
The holographic technique is composed of four steps as following.

First, before the atomic imaging measurements, we pre-characterize the wavefront of the probe field, $E_{\rm p}$, through multi-plane intensity measurements followed by a multi-plane Gerchberg-Saxton (G-S) algorithm to reconstruct the phase information~\cite{Gerchberg1972,Ivanov1992}. 

Next, for the holographic imaging, defocused images $I_1=|E_{\rm p}+E_{\rm s}|^2$ and $I_2=|E_{\rm p}|^2$ are recorded at a ``holography'' plane $z_{\rm H}$, with and without the atomic sample respectively, as outlined in the main text (Fig. 1). Taking into account ambient noise that affect both $I_1$ and $I_2$, in this step a numerical procedure is applied to optimally match $I_2$ to  the  pre-characterized $E_{\rm p}$ and to derive the reduced hologram $\delta I=I_1-I_2$.

Thirdly, $E_{\rm s}$ is numerically reconstructed from the reduced hologram $\delta I=I_1-I_2$ with an iterative algorithm to remove the twin and dc noises~\cite{Sobol2014a}.

Finally, $E_{\rm s}$ and $E_{\rm p}$ are numerically  propagated to atomic plane $z=z_{\rm A}$ to evaluate the complex phase $\varphi(x,y)=\phi+i {\rm OD}/2=-i {\rm log}(\frac{E_{\rm p}(z_{\rm A})+E_{\rm s}(z_{\rm A})}{E_{\rm p}(z_{\rm A})})$. 

As in the main text, we may omit the $x,y$ (and $k_x,k_y$) coordinate variables in the notions of wavefronts and intensities if ambiguity would not be introduced.

To propagate a wavefront $E(x,y,z)$ from $z=z_A$ to $z=z_B$, we use the angular spectrum method with the following notations:
\begin{equation}
\begin{array}{l}
E(z_B)=\hat U(z_B-z_A)E(z_A),{\rm with}\\
\hat{U}(L)={\hat{F}}^{-1}e^{i  \sqrt{k^2-k_x^2-k_y^2}L}\hat{F}
\end{array}\label{eq:asm}
\end{equation}
Here $k=2\pi/\lambda$ is the wavenumber of the light field, and $\hat{F}$ represent the 2D Fourier transform~\cite{kim2010principles}, $i.e.$, $\hat F E(x,y,z)=E(k_x,k_y,z)$ and $\hat F^{-1} E(k_x,k_y,z)=E(x,y,z)$.

In the following we detail the four-step holographic reconstruction procedure.

\subsection{Pre-characterization of $E_{\rm p}$}

\begin{figure}[htbp]
    \centering
    \includegraphics[width=\linewidth]{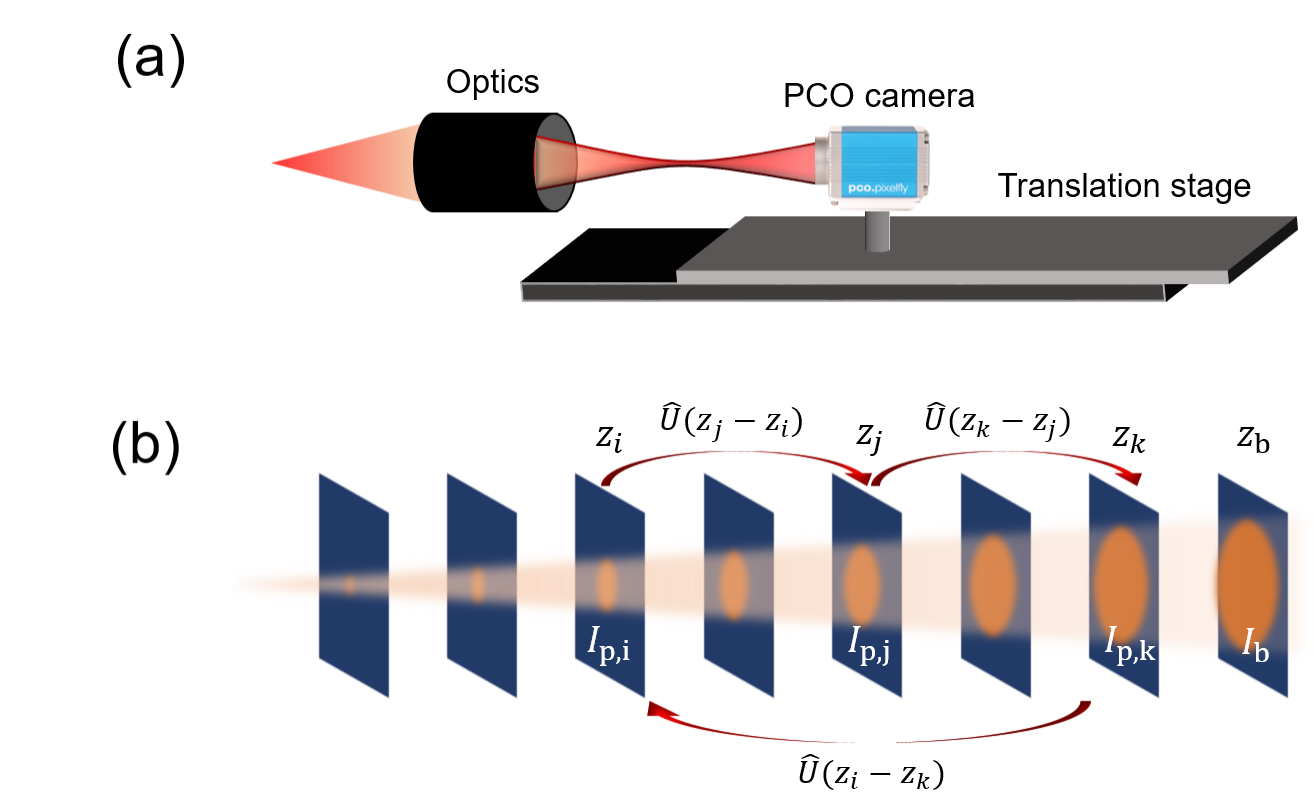}
    \caption{Schematic setup for the measurement (a) and numerical reconstruction (b) of the probe light wavefront $E_{\rm p}$. The intensity profile $I_{\rm p}(z_j)=|E_{\rm p}(z_j)|^2$ is recorded at multple $z-$planes with a digital camera mounted on a translation stage. The flow diagram in (b)  describes a single 3-plane G-S iteration for the complex $E_{\rm p}$ reconstruction. The full reconstruction is composed of multiple 3-plane G-S iterations with $\{I_{\rm p,j}\}$ data randomly chosen from the pre-recorded data-set. }
    \label{FigS-GS1} 
\end{figure}

As in Fig.~\ref{FigS-GS1}(a), we use a CCD camera mounted on a stepper-motor-driven linear stage to sample the probe field intensity distribution $I_p(x,y,z)==|E_{\rm p}(x,y,z)|^2$ in 3D, by translating the camera along $z$ and record a set of 2D intensity profiles $I_{\rm p,j}=I_{\rm p}(z_j)$. Here the $x-y$ plane is defined by the camera sensor chip. The direction normal to the plane is thus referred to as the $z-$direction, which is adjusted to be parallel to the direction of translation to within $\sim$10~mrad angular precision. At each $z_j$-plane, the intensity $I_j$ is adjusted with an acoustic-optical modulator (AOM) to ensure sufficient counts on the camera without saturation. The distance $z_j$ is digitally recorded, which is subjected to numerical adjustments as detailed in the following. The imaging data is referred to as a $\{I_{\rm p}\}$-set. 

The 3D intensity sampling is subjected to systemmatic noises such as multiple reflections involving the camera sensor chip itself. To recover $E_{\rm p}$ from the imperfect $\{I_{\rm p}\}$ measurement, it is important to sample many $z_j-$ planes to suppress the position-dependent noise so as to determine $E_{\rm p}$ through the average intensity field.  To achieve the purpose without demanding computer memory, we apply a multi-plane G-S algorithm that samples the $\{I_{\rm p}\}$-set in small groups, detailed as following. It should be noted that although we take a strongly focused probe beam (NA=0.15 after all the imaging optics) as the example in this work, the method is applicable to reconstructing general form of probe beam wavefronts.

\begin{enumerate}
\item[(1)] Split pixels: each image of the camera-recorded $I_p$-set, as represented by a $M\times N$ 2D matrix, is split by a factor of $n_s\times n_s$ to enhance the spatial resolution $\Delta x=\Delta x^{(0)}/n_s$. Here $\Delta x^{(0)}$ is the physical camera sensor resolution. For the CCD camera in this work, with $M=1240$, $N=1392$, and $\Delta x^{(0)}=6.45~\mu$m, we find $n_s=4$ is large enough to support the complex $E_{\rm p}$ reconstruction at $NA=0.15$. We choose $n_s=6$ with large enough margin in this work.

\item[(2)] Set a ``bank'' plane: We choose a ``bank'' plane $z_{\rm b}$ and the associated probe beam intensity profile $I_{\rm b}$ from the $\{I_{\rm p}\}$-set. The phase of the wavefront $E_{\rm b}$ is initialized as a spherical wave phase $\phi_{\rm p}^{\rm s w}=k\sqrt{x^2+y^2+z_{\rm b}^2}$. To do this, we have assumed $z=0$ for the $E_{\rm p}$ focal plane, which is estimated by additional camera images near $z=0$. An initial guess of $E_{\rm p}$ as $E_{\rm p}^{G S} = \sqrt{I_{\rm b}} e^{i\phi_{\rm p}^{\rm s w}}$ starts the G-S iteration in the next step. With $z_{\rm b}$ chosen as the ``bank'' plane, $I_{\rm b}$ is subsequently removed from the $\{I_{\rm p}\}$ data-set.

\item[(3)] Three images from the $\{I_{\rm p}\}$-set are randomly chosen with intensity distribution $I_{\rm p,i},I_{\rm p,j},I_{\rm p,k}$ recorded at $z_i,z_j,z_k$.  With the angular spectral method by Eq.~(\ref{eq:asm}), $E_{\rm p}^{GS}$ is first propagated to $z_i$ as $E_{\rm p}^{GS} \leftarrow \hat{U}(z_i-z_{\rm b}) E_{\rm p}^{GS}$.

\item[(4)] The following G-S iteration among the $i,j,k$ planes are applied  for $n_a$ times as following (Fig.~\ref{FigS-GS1}b): Cyclically propagate $E_{\rm p}^{GS}$ from $z_i$ to $z_j$, $z_k $ and back to $z_i$. In each step and before the next numerical propagation,  the amplitude of $E_{\rm p}^{GS}$ is replaced with the camera-recorded $\sqrt{I_{\rm p}}$ while keeping the phase information. We usually set $n_a=30$.

\item[(5)] 
With the $z_{i,j,k}$ values initially estimated from the linear stage readouts, the $z_{i,j,k}$ optimization procedures are sparsely inserted into the G-S iteration in step 4. For example, when $E_{\rm p}^{GS}$ is propagated from $z_i$ to $z_j$, the value $z_j$ is adjusted so as to optimize the inner product between the $|E_{\rm p}^{GS}(z_j)|^2$ and $I_{\rm p, j}$ images. The corrections to $z_{i,j,k}$ are updated in the following iterations.

\item[(6)] Add to the ``bank'' wavefront: Propagate $E_{\rm p}^{GS}$ back to the ``bank'' plane $z_{\rm b}$. Add $E_{\rm p}^{GS}$ to the ``bank'' waveform $E_{\rm p}^{Bank}$ with an exponential weight: 
$E_{\rm p}^{Bank} \leftarrow \frac{1}{1+\eta} ( E_{\rm p}^{Bank} + \eta E_{\rm p}^{GS})$. Here $\eta$ is a weight factor that determine the convergence speed and stability. After the bank-addition procedure, the G-S wavefront for the iteration is updated with the best estimated value at the ``bank'' location: $E_{\rm p}^{GS} \leftarrow E_{\rm p}^{\rm Bank}$.

\item[(7)] Repeat the (3)-(6) loop $n_b$ times until the reconstruction similarity $s=\langle \delta I_{\rm t}|\delta |E_{\rm p}^{GS}(z_{\rm t})|^2\rangle$, an ``inner product'' between the reconstructed $|E_{\rm p}^{GS}(z_{\rm t})|^2$ and $I_{\rm t}$ profiles converge. Here $I_{\rm t}$ the intensity distribution at a target plane far away from the planes for the $\{I_p\}$ set. The reconstructed reference wavefront is finally obtained at $z=z_{\rm b}$ as $E_{\rm p} = E_{\rm p}^{GS}$, and is now free to propagate to any other $z$-plane.

\end{enumerate}

Here, with $\delta I_A=I_A-\bar I_A$ and $\delta I_B=I_B-\bar I_B$ the similarity is defined as
\begin{equation}
    \langle \delta I_A|\delta I_B\rangle=\frac{\overline{\delta I_A \delta I_B}}{\sqrt{\overline{\delta I_A^2}}\sqrt{\overline{\delta I_B^2}}}.
\end{equation}

\begin{figure}[htbp]
    \centering
    \includegraphics[width=\linewidth]{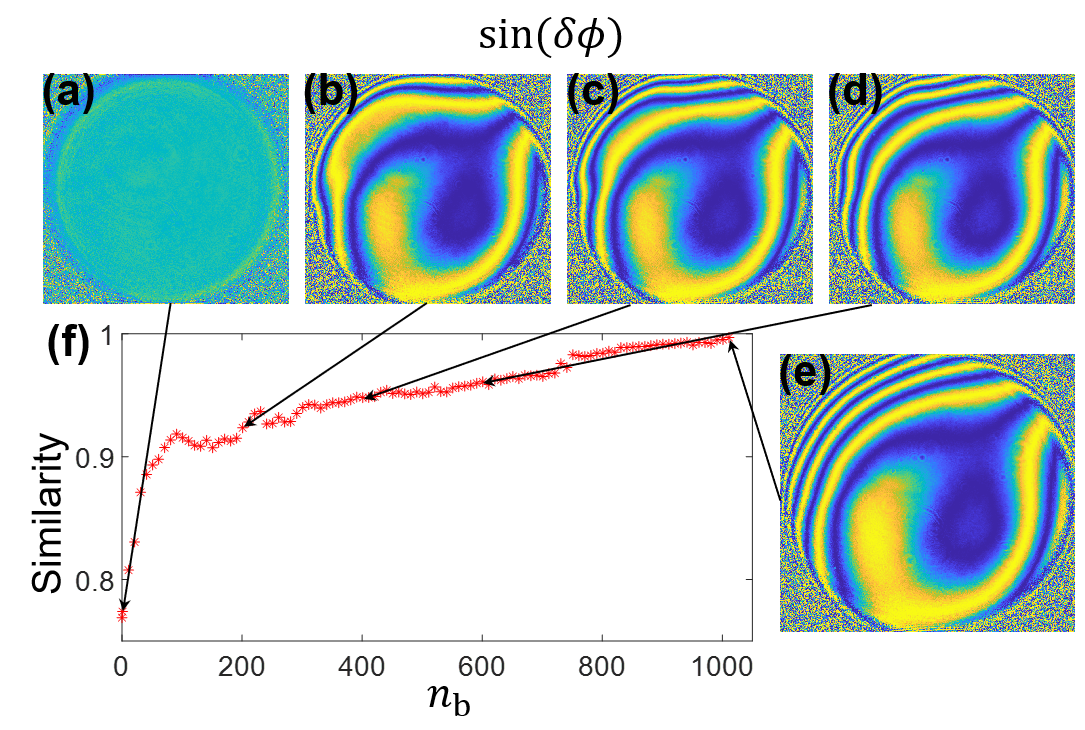}
    \caption{Evolution of the reconstruction fidelity $f$. Typical phase difference between iterated wavefront and the spherical wave $\delta\phi_{\rm p}=\phi_{\rm p}-\phi_{\rm p}^{\rm s w}$ (Fig. a-e) are also presented. The final wavefront }
    \label{FigS1-1} 
\end{figure}

In Fig.~\ref{FigS1-1} we plot the evolution of the similarity $s$ during a typical $E_{\rm p}$ reconstruction, together with the reconstructed wavefront phase $\phi_{\rm p}$ at typical steps of the G-S iteration. The G-S process typically take $n_b=1000$ steps toward full convergence. The final intensity similarity $s\approx 0.99$ suggests a similar level of fidelity for the reconstructed wavefront $E_{\rm p}=\sqrt{I_{\rm p}}e^{i\phi_p}$. The slow iteration is a result of low-efficiency correction to the large scale wavefront aberrations. In ref.~\cite{futurework} an optimization procedure that substantially speeds up the process, as well as a full characterization of the wavefront recosntruction fidelity will be discussed.

\subsection{Hologram recording, reduction and $E_{\rm p}$ matching\label{sec:Subtraction}}

In this step, holograms of atomic sample are recorded as $I_1$ and $I_2$, with and without the atomic sample respectively as outlined in the main text (Fig.~1). To suppress the impact of vibration noises during the $I_{1,2}$ recording, we minimize the inter-frame time to be merely 10~ms just to allow the atomic sample to escape the probe area. We then optimally subtract away the ambient background recorded by additional background measurements, following a procedure detailed in ref.~\cite{Sobol2014}. The procedure allows us to obtain the reduced hologram $\delta I=|E_{\rm p}+E_{\rm s}|^2-|E_{\rm p}|^2$ for the $E_{\rm s}$ retrieval in the next step. To account for drifts of optics alignments, We perform a 3D numerical shift of $E_{\rm p}$ to optimally match the pre-characterized $|E_{\rm p}(z_{\rm H})|^2$ with $I_2$.

\subsection{$E_{\rm s}$ retrieval}

We now divide the reduced hologram $\delta I$ with the optimally-matched $E_{\rm p}^*$ to obtain our initial guess of $E_{\rm s}$,
\begin{equation}
E_{\rm diff}=\frac{\delta I}{{E_{\rm p}}^*}=E_{\rm s}+\frac{{E_{\rm s}}^*E_{\rm p}}{{E_{\rm p}}^*}+\frac{|E_{\rm s}|^2}{{E_{\rm p}}^*}
\label{Eq-2-2}
\end{equation}


\begin{figure}[htbp]
    \centering
    \includegraphics[width=\linewidth]{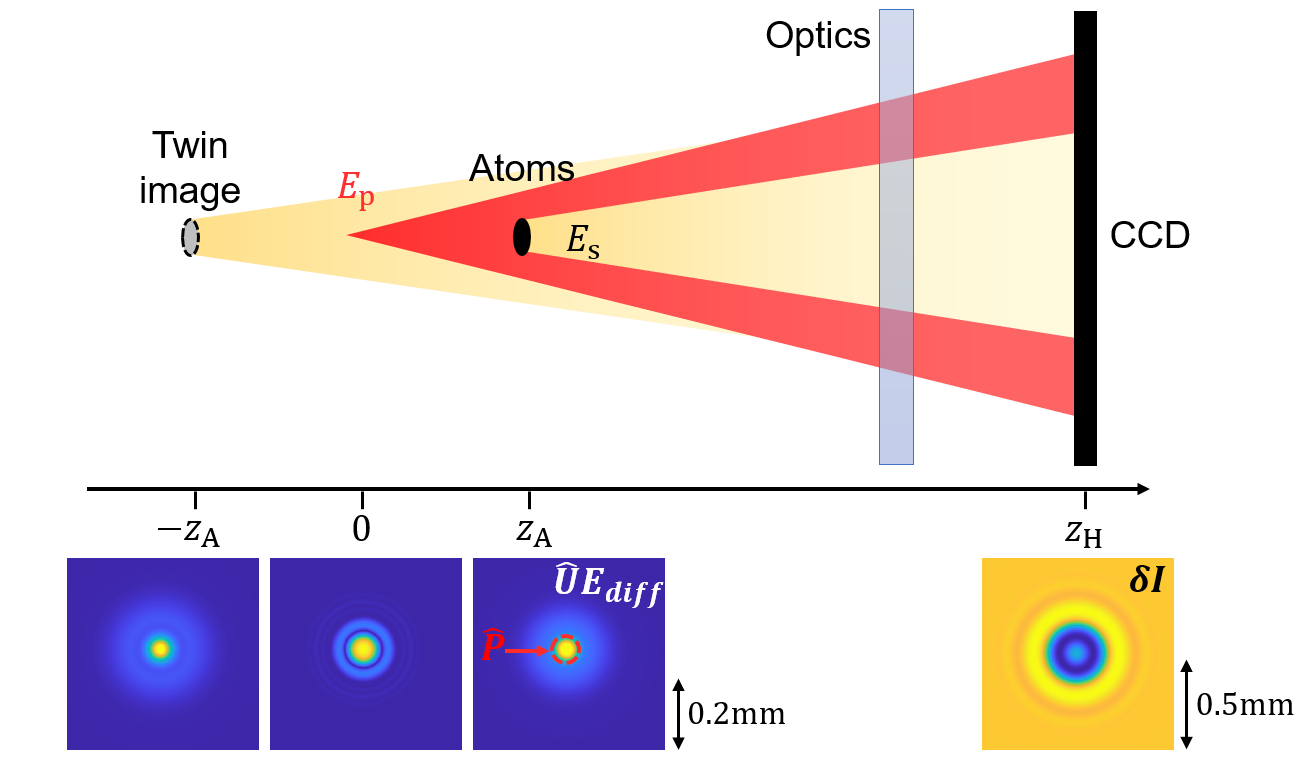}
    \caption{Schematic illustration of the $E_{\rm s}$, twin image, DC term. Here $E_s$ is focused at atomic plane $z=z_{\rm A}$. DC and twin images are focused approximately at $z=0$ and $z=-z_{\rm A}$ respectively. To remove the twin and dc noise, a $\hat{P}$ operation can be introduced by picking up the $E_{\rm s}$ signal with concrete support.}
    \label{FigS2-1} 
\end{figure}
As illustrated in Fig.~\ref{FigS2-1}, $E_{\rm diff}$ contains information of $E_{\rm s}$ that properly focuses at the atomic sample plane $z=z_{\rm A}$. In addition, the twin-image term $E_{\rm s}^*E_{\rm p}/E_{\rm p}^*$ focuses approximately at $z=-z_{\rm A}$, while the dc noise $|E_{\rm s}|^2/E_{\rm p}^*$ focuses at $z=0$. Due to the inline geometry, these additional noise terms overlap with $E_{\rm s}$ and there is no simple method to isolate $E_{\rm s}$ without a prior knowledge. Here we employ minimal amount of a prior knowledge on the likely location of the atomic sample, and use an iterative algorithm~\cite{Sobol2014} to remove the twin and dc noises and to faithfully retrieve $E_{\rm s}$ from $E_{\rm diff}$, as following:

First, we define an aperture operator $\hat P$, which sets $\hat P E(z_{\rm A})=E(z_{\rm A})$ inside the aperture and $\hat P E(z_{\rm A})=0$ outside the aperture (Fig.~\ref{FigS2-1}). With the concrete support of the $E_s$ signal by the a prior location information, $\hat P E_{\rm s}(z_A)=E_{\rm s}(z_A)$,  $\hat P E_{\rm diff}(z_A)$ picks up all the $E_{\rm s}$ but removes part of the twin and dc noises. To iteratively improve the isolation of the $E_{\rm s}$, we take advantage of the symmetry between $E_{\rm s}$ and the twin-image terms to introduce a complex conjugation operator $\hat C$ at $z=z_{\rm H}$ plane as~\cite{Sobol2014}, 

\begin{equation}
\hat{C}E(z_{\rm H}) = \frac{E^*(z_{\rm H})E_{\rm p}(z_{\rm H})}{E_{\rm p}^*(z_{\rm H})}.
\label{Eq-2-4}
\end{equation}

A iteration procedure then follows
\begin{equation}
E_{\rm s}^{(n+1)} = \hat{U}\hat{C}(E_{\rm diff}-\hat{U}^{-1}\hat{P}E_{\rm s}^{(n)})
\label{Eq-2-3}
\end{equation}
to remove the twin and dc noise.

The initial value of of the $E_{\rm s}^{(n)}$ is set as $E_{\rm s}^{(n=0)} = \hat{U}E_{\rm diff}$. As discussed in ref.~\cite{Sobol2014,Sobol2014a}, $E_{\rm s}^{(n)}$ converges to $E_{\rm s}$ with a speed determined by the fraction of light field energy removed by $\hat P$ at each step.  Unlike the work in ref~\cite{Sobol2014} where a complex $E_{\rm p}$ composed of two point sources are introduced, here we simply use a nearly spherical $E_{\rm p}$. Fast convergence of the iteration is guaranteed by a small atomic sample with size $\sigma^2\ll \lambda z_A$ to support a relatively small $\hat P$ $E_{\rm s}$ aperture. On the other hand, to image large atomic sample with $\sigma^2\sim \lambda z_{\rm A}$ we should increase the $E_{\rm p}$ complexity accordingly as those in ref.~\cite{Sobol2014}.

Finally, after the convergence of iteration by Eq.~(\ref{Eq-2-3}), the removal of the dc noise is achieved by updating the iteration relation as~\cite{Sobol2014}

\begin{equation}
E_{\rm s}^{(n+1)} = \hat{U}\hat{C}(E_{\rm diff}-\hat{U}^{-1}\hat{P}E_{\rm s}^{(n)} - \frac{|\hat{U}^{-1}\hat{P}E_{\rm s}^{(n)}|^2}{{E_{\rm p}}^*}).
\label{Eq-2-5}
\end{equation}

\begin{figure}[htbp]
    \centering
    \includegraphics[width=1\linewidth]{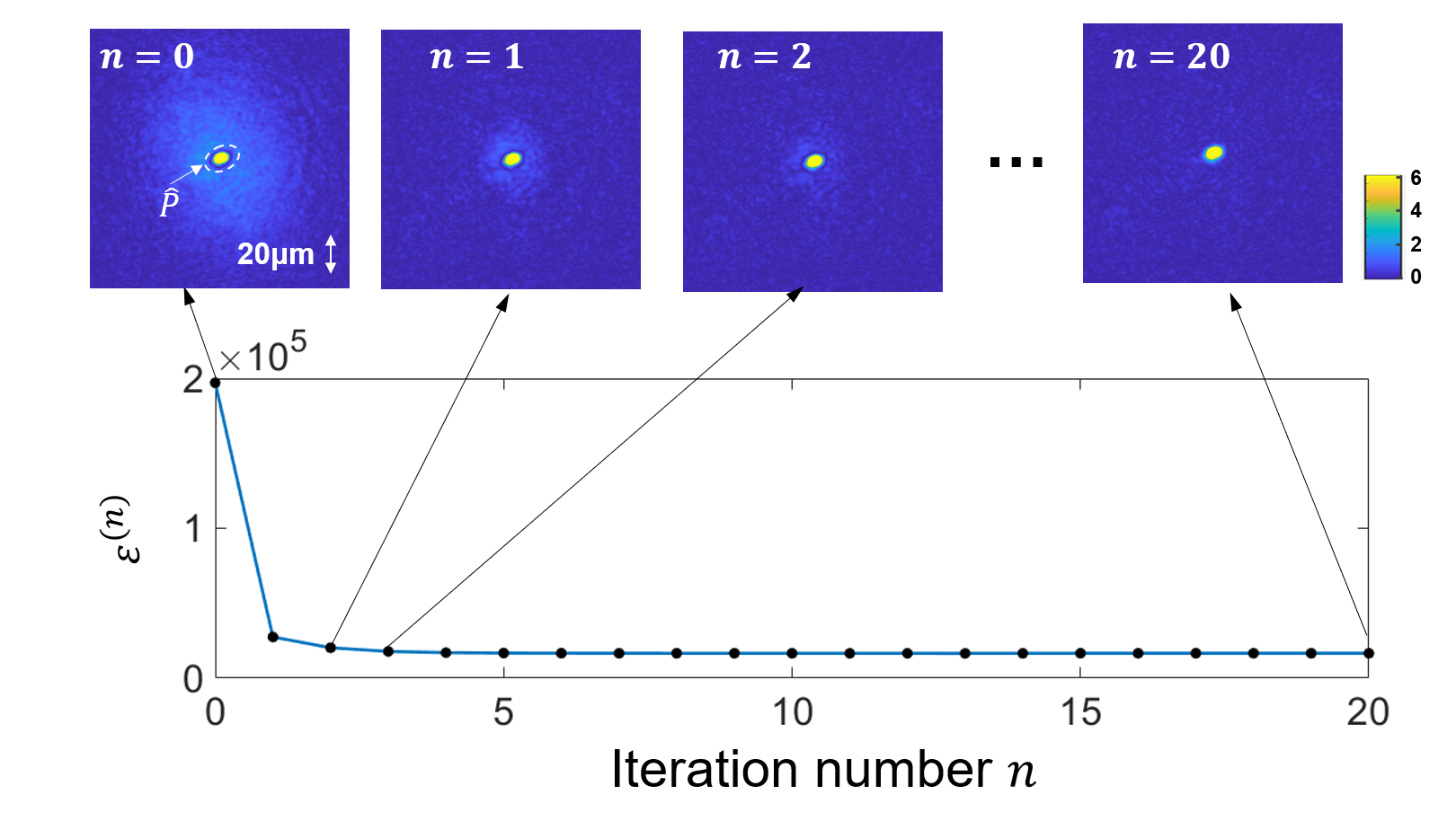}
    \caption{Convergence of the residual energy     
    $\varepsilon^{(n)}$ during the twin-DC removal iteration. Here $\varepsilon^{(n)}$ is summed over the displayed imaging area on the top, with the $\hat P$ area excluded.}
    \label{FigS2-2} 
\end{figure}

We characterize the convergence of iteration with a residual energy, $\varepsilon^{(n)}=\sum |(1-\hat P)E_{\rm s}^{(n)}|^2$ summing over an area substantially larger than the sample size at $z_A$. Typical convergence of the residual is shown in Fig.~\ref{FigS2-2}.
Importantly, this algorithm does not assume any special form of reference wavefront $E_{\rm p}$, nor paraaxial approximations that limit the spatial frequency of the wavefront.

\subsection{${\rm OD}$ and $\phi$ retrieval}

In this last step, we propagate the reconstructed $E_{\rm p}$ and $E_{\rm s}$ to the plane of atomic sample, $z=z_{\rm A}$, to obtain $\rm OD$ and $\phi$ data as the imaginary and real part of complex phase $\varphi(x,y)=-i {\rm log}(\frac{E_{\rm p}+E_{\rm s}}{E_{\rm p}})$. Here, with full knowledge of $E_{\rm p}$ and $E_{\rm s}$, various coherent imaging-processing steps can be taken to suppress imaging aberration and speckle noises. For example, to reduce the impact of speckle noises in $E_{\rm p}$ which are introduced by the imaging system itself (not seen by the atoms), we first propagate $E_{\rm p}$ to $z=0$ to limit its value within the spread of $E_{\rm s}$, before propagating the filtered $E_{\rm p}$ back to the $z=z_{\rm A}$ plane for the atomic response retrieval. It is worth noting these speckle noises should equally affect signals in regular absorption images where a similar correction is difficult to make.

\section{Details on data analysis}

\subsection{Fitting the complex EIT spectrum}\label{sec:EIT}
\begin{figure}[htbp]
    \centering
    \includegraphics[width=0.5\linewidth]{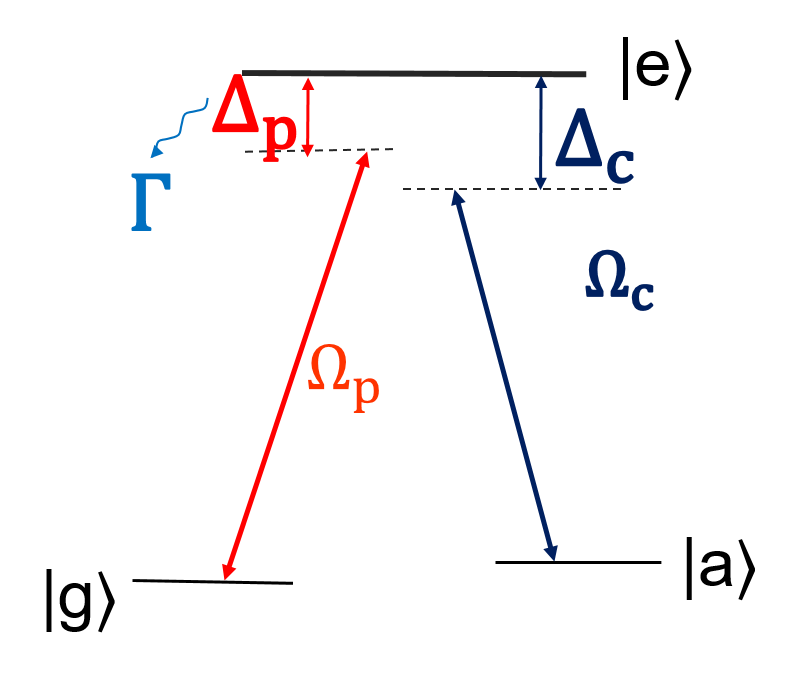}
    \caption{Schematic of a 3-level system system to obtain the phenomelogical EIT response.}
    \label{fig:3level} 
\end{figure}

The complex spectrum obtained from the holographic measurements can be fit  to a phenomenological EIT model to assist the data analysis. The fit process is detailed in this section. Here, instead of a full analysis of EIT for the multi-level atom in presence of residual magnetic field,  we use a simple 3-level model (Fig.~\ref{fig:3level}) and perform a linear analysis of the atomic response to $E_p$, leading to the standard expression of the atomic polarizability,

\begin{equation}
    \alpha(\Delta_\mathrm{p},\Delta_\mathrm{c})=\alpha_0+\frac{C}{\frac{|\Omega_\mathrm{c}|^2}{4(\Delta_\mathrm{p}-\Delta_\mathrm{c}-i\kappa/2)}-(\Delta_p+i\Gamma/2) }.
  \label{eq:alpha-probe-2}
  \end{equation}
Here the parameter $C$  determines the $|g\rangle \leftrightarrow |e\rangle$ effective transition strength. The $\Delta_\mathrm{p}=\omega_\mathrm{p}-\omega_\mathrm{eg}$ and $\Delta_\mathrm{c}=\omega_\mathrm{c}-\omega_\mathrm{ea}$ are the probe and control laser detunings respectively. $\kappa$ is a phenomenological damping constant to account for relaxation between the ground state $|g\rangle $ and $|a\rangle$. A non-zero $\alpha_0$ is expected from real atomic response due to non-ideal optical pumping, additional magnetic coupling, as well as off-resonant coupling of $E_\mathrm{p}$ to additional hyperfine transitions.

For the dilute and optically thin atomic sample in this work, the  complex phase shift $\varphi(\omega_\mathrm{p},\omega_\mathrm{c})$ obtained from the holography is expected to be proportional to $ \alpha(\omega_\mathrm{p},\omega_\mathrm{c})$. We thus apply  Eq.~(\ref{eq:alpha-probe-2}) to fit the experimental complex phase from the $\mathrm{OD}$ and $\phi$ data in Figs. 2, 3, 4 of the main text for the case of stationary atoms. The free parameters $\alpha_0, C, |\Omega_\mathrm{c}|^2, \kappa, \Gamma, \Delta_\mathrm{c}$ for the Fig. 2 fit are given by: $\alpha_0=0.06-0.141i$, $C=4.06$, $\Delta_\mathrm{c}/(2\pi)=-0.06$MHz, $\Gamma=12.03/(2\pi)$~MHz, $|\Omega_\mathrm{c}|^2/(2\pi)^2=26.06$~MHz$^2$, $\kappa/(2\pi)=0.56$~MHz.



\subsection{Analysis of imaging noise}

\begin{figure*}[htbp]
    \centering
    \includegraphics[width=0.85 \textwidth]{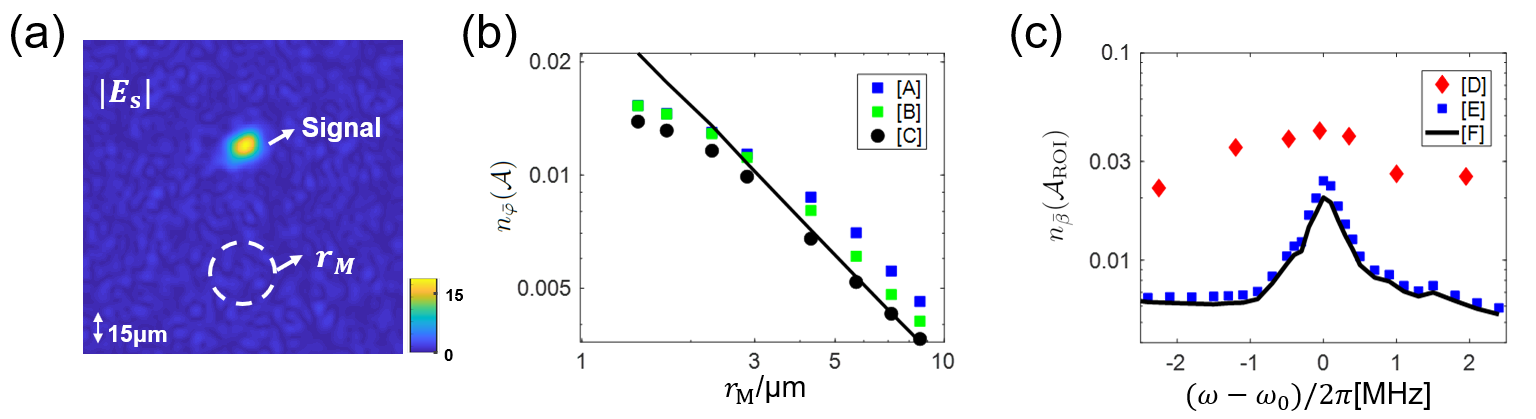}
    \caption{\label{fig:noiseF} Figure~3 of the main text. See the text here for discussions.}
  \end{figure*}

This section provides more details on imaging noise analysis for Fig.~2 of the main text, which is duplicated here as Fig.~\ref{fig:noiseF}.

Similar to absorption and phase contrast imaging techniques, noises in holographic imaging is fundamentally limited to the photon shot noise of the probe light~\cite{Sobol2014}. In this work, atomic properties are retrieved from the noisy ${\rm OD}(x,y)$ and $\phi(x,y)$ images reconstructed from the holograms. The signal retrieval is defined within a region of interest ${\rm ROI}$ with area $\mathcal{A}_{\rm ROI}$ that include the atomic signals. To evaluate the noise level associated with the measurements, one could repeat the same measurements many times to look for statistical fluctuations. However, this approach is influenced by various technical fluctuations irrelevent to the imaging technique itself. These include laser intensity and frequency noise and fluctuation of atom number during repeated measurements. As clearly illustrated in Fig.~2c of the main text, for  $\overline{\rm OD}$ and $\bar\phi$ 
measuements in this work, the dominant channel of technique fluctuation is atom number fluctuation.

To recover intrinsic sensitivity of the measurements in presence of such frequency fluctuation,  we take the advantage of the fact that the holographic imaging processing is essentially a linear process (except the dc removal part by Eq.~(\ref{Eq-2-5}), whose impact is typically negligible for weak $E_{\rm s}$ signals), with the noise level to ${\rm OD}$ and $\phi$ not affected by the presence of the atomic signal (Fig.~\ref{fig:noiseF}a). We therefore analyze the statistical noise level within area $\mathcal{A}$ for the following three cases (Fig.~\ref{fig:noiseF}b), and infer from the agreed noise level as that for the shot-noise-limited measurements for the atomic signals. The three cases are:

\begin{itemize}
\item{[A]:} {Experimental data within $\mathcal{A}$ excluding the atomic signals ((Fig.~\ref{fig:noiseF}a). } 

\item{[B]:} {Experimental data measured in absence of atomic sample, within the same area $\mathcal{A}$ .} 

\item{[C]:} {Simulated holographic data with shot-noise level  estimated from the experimentally measured intensity $I_{1,2}$, assuming the CCD camera counts following a fluctuation with poissonian statistics. }
\end{itemize}

The results of the noise analysis are plotted in Fig.~\ref{fig:noiseF}b. Here we consider the setup geometry in this work, with a large distance $|z_H-z_A|$ between the camera and atomic plane. In this case, for small $\mathcal{A}=\pi r_M^2$ close to or smaller than $\mathcal{A}_{\rm ROI}$ (corresponding to $r_M\approx 10~\mu$m) we find the noise level is isotropic on the complex plane with $n_{\bar \phi}\approx n_{\overline{\rm OD}/2}$, and therefore the total noise level is given by $n_{\bar \varphi}\approx \sqrt{2}n_{\bar \phi}$. Furthermore, with the noise level from the shot-noise limited simulation (case C) nicely follow the curve $n_{\bar\varphi}(A)=1/\sqrt{N_p(\mathcal{A})}$, the experimentally estimated noise levels with (case A) and without (case B) atoms  are typically a factor of $20-50\%$ beyond the shot-noise limit, likely caused by imperfect camera readouts and $I_{1,2}$ subtractions.

Experimental noise is estimated in an area $\mathcal{A}$ where atom signal is excluded(see Fig. \ref{fig:noiseF}(a)). The center of mask is about 30$\mu$m from the atom, well within the center of imaging area, which guarantees its same noise features with atomic area. The noise is interpreted a summation of $\mathrm{OD}/2$ and $\phi$ noise:

\begin{equation}   n_\mathrm{\bar\varphi}=\sqrt{(\frac{1}{2}n_{\overline{\mathrm{OD}}})^2+n_{\bar\phi}^2}  \label{eq:s-noise} \end{equation}

The two terms on the right hand is the standard deviation calculated from 100 pictures for each case in Fig. \ref{fig:noiseF}(b). To make out the spectrum of noise, the radius $r_\mathrm{M}$ of $\mathcal{A}$ is scanned. Case [C] shown by the red spots contains only shot noise, and matches well with the $y=1/\sqrt{N_\mathrm{p}(A)}$ curve at big $r_{\rm M}$ . Noise for the other two cases are 20\% $\sim$ 50\% larger.

Also in Fig.~\ref{fig:noiseF}, it is important to point out that when $r_{\rm M}$ is smaller than resolution, the noise inside $\mathcal{A}$ has correlation and $n_{\bar \varphi}(A)$ deviates from the $1/\sqrt{N_p(\mathcal{A})}$ power law. The turning point of the noise spectrum corresponds to the size of the imaging resolution $\delta x$. Here, since during the holography reconstruction process we perform a spatial filter for both $E_{\rm r}$ and $E_{\rm s}$ on the holography plane, with a 0.7~mm radius mask, the effective numerical aperture of the final images is reduced from $NA=0.45$ to 0.13, leading to $\delta x\approx 6\mu$m spacial resolution and the deviation in Fig.~\ref{fig:noiseF}b.

On the other hand, since the phase angle $\beta$ measurement is immune to atom number fluctuation, we have the opportunity to evaluate shot-to-shot fluctuation of $\bar \beta(\mathcal{A}_{\rm ROI})$ and to compare it with theoretical shot-noise-limit.

\begin{itemize}
    \item{[D]:} {. } 
    
    \item{[E]:} { .} 
    
    \item{[F]:} {. }
    \end{itemize}


\subsection{Inference of $\rho_c-u$ distribution}

A key advantage of coherent imaging is the ability to simultaneous retrieve $OD$ and $\phi$ images from single-shot holograms. For a dilute and optically thin gas, this allows us to infer atomic density $\rho_c$ and single atom polarizability $\alpha$ from the complex $\varphi=\phi+i{\rm OD}/2$ through $\varphi=k \rho_c \alpha$. This section describes the procedure to infer atomic density $\rho_c$ and velocity field $u$ from the reconstructed $\varphi$ data,  using the EIT response as described in Sec.~\ref{sec:EIT}.

\bibliography{sl}